\documentclass[a4paper,12pt]{article}

\usepackage{epsf, amssymb,amsthm, amsmath,dsfont}
\usepackage{epsfig}
\usepackage{hyperref}
\usepackage{xcolor}
\usepackage{graphicx} 
\usepackage{comment}
\usepackage{cite}

\counterwithin*{equation}{section}

\hypersetup{hidelinks} 

\makeatletter
\renewcommand\section{\@startsection {section}{1}{\z@}%
                                   {-3.5ex \@plus -1ex \@minus -.2ex}
                                   {2.3ex \@plus.2ex}%
                                   {\normalfont\large\bfseries}}

\renewcommand\subsection{\@startsection{subsection}{2}{\z@}%
                                     {-3.25ex\@plus -1ex \@minus -.2ex}%
                                     {1.5ex \@plus .2ex}%
                                     {\normalfont\bfseries}}

\makeatother

\newcommand{\bea}{\begin{eqnarray}}
\newcommand{\eea}{\end{eqnarray}}
\newcommand{\be}{\begin{equation}}
\newcommand{\ee}{\end{equation}}
\newcommand{\bem}{\begin{pmatrix}}
\newcommand{\eem}{\end{pmatrix}}
\newcommand{\bl}{\begin{align}}
\newcommand{\el}{\end{align}}

\newcommand{\beq}{\begin{equation}}
\newcommand{\eeq}{\end{equation}}
\newcommand{\tr}{\mathrm{tr}}
\newcommand{\cM}{\mathcal{M}}
\newcommand{\cN}{\mathcal{N}}
\newcommand{\cH}{\mathcal{H}}

\def\i{\iota}

\def\cm{{\cal M}}
\def\cn{{\cal N}}

\newtheorem{theorem}{Theorem}[section]

\newtheorem{proposition}[theorem]{Proposition}
\theoremstyle{definition}
\newtheorem{definition}{Definition}[section]
\theoremstyle{remark}

\begin{document}

\begin{center} 
${}$
\thispagestyle{empty}

\vskip 1.5cm {\Large {\bf An Operator Algebraic Approach \\[3mm] To Black Hole Information}}
\vskip 1cm  { Jeremy van der Heijden\footnote{
		j.j.vanderheijden2@uva.nl}\\[3mm]
Erik Verlinde\footnote{
e.p.verlinde@uva.nl} }\\
{\vskip 0.5cm  Institute for Theoretical Physics\\ University of Amsterdam\\
Science Park 904\\
1090 GL Amsterdam\\ 
The Netherlands}

\vspace{1cm}

\begin{abstract}
\baselineskip=16pt

We present an operator algebraic perspective on the black hole information problem. For a black hole after Page time that is entangled with the early radiation we formulate a version of the information puzzle that is well-posed in the $G\to 0$ limit. We then give a description of the information recovery protocol in terms of von Neumann algebras using elements of the Jones index theory of type II$_1$ subfactors. The subsequent evaporation and recovery steps are represented by Jones's basic construction, and an operation called the canonical shift. A central element in our description is the Jones projection, which leads to an entanglement swap and implements an operator algebraic version of a quantum teleportation protocol. These aspects are further elaborated on in a microscopic model based on type I algebras. Finally, we argue that in the emergent type III algebra the canonical shift may be interpreted as a spacetime translation and, hence, that at the microscopic level “translation = teleportation”. 

\end{abstract}

\end{center}
\newpage

\tableofcontents

\section{Introduction}
The study of operator algebras has gained considerable attention recently in the context of black hole physics and the AdS/CFT correspondence. It was shown that the boundary CFT associated to the eternal AdS black hole spacetime exhibits an emergent type III$_1$ von Neumann algebra factor in the infinite $N$ limit \cite{Leutheusser:2021qhd, Leutheusser:2021frk}. Furthermore, it was argued that, in this setting, by including $1/N$ corrections the algebra becomes a type II$_\infty$ von Neumann algebra which is related to the type III$_1$ algebra via the so-called crossed product construction \cite{Witten:2021unn}. Although the mathematical machinery is by no means new, what makes their recent application stand out is the specific focus on the role of gravity and the emergence of spacetime, see, for example \cite{DeBoer:2019kdj, Chandrasekaran:2022cip, Chandrasekaran:2022eqq, deBoer:2022zps, Furuya:2023fei, Jensen:2023yxy, AliAhmad:2023etg, Gesteau:2023rrx, Ouseph:2023juq, Engelhardt:2023xer, Gesteau:2024dhj, Akers:2024bel, Faulkner:2024gst}.

In the present paper, we study the application of operator algebras to the problem of black hole information. An important step towards resolving the information paradox was made by the quantum extremal surface (QES) prescription \cite{Engelhardt:2014gca} and the corresponding notion of entanglement islands \cite{Penington:2019npb, Almheiri:2019psf, Almheiri:2019hni}. These semi-classical methods confirm essential features of a unitary microscopic quantum description, in particular the fact that the entanglement entropy of the radiation follows the Page curve. To fully resolve the information paradox, however, it is necessary to derive this emergent spacetime description from the underlying microscopic theory. For example, it is still not known how to describe the experience of the in-falling observer in the {\it same} microscopic theory that explains the retrieval of the quantum state of the in-falling matter. In addition, one would like to derive the QES prescription and appearance of entanglement islands from this microscopic description.

Arguably, the origin of the information puzzle can be traced back to differences in the type\footnote{For some background on the type classification of von Neumann algebras from a physical perspective we refer to, e.g., \cite{Witten:2018zxz, Sorce:2023fdx}.} of operator algebras that are required for the microscopic and emergent spacetime descriptions. In the microscopic theory, an evaporating black hole is expected to behave as an ordinary quantum system with a finite number of states given by the Bekenstein-Hawking entropy. In such a theory, the evaporation process is unitary and hence, the quantum information contained in the black hole states will eventually be transferred to the outside radiation. Furthermore, starting with the work of Hayden and Preskill \cite{Hayden:2007cs} it is by now quite well understood how to think about the information recovery process and its relation to the methods of quantum error correction \cite{Verlinde:2012cy, Almheiri:2014lwa, Dong:2016eik, Harlow:2016vwg, Yoshida:2017non, Akers:2022qdl}. On the level of operator algebras, the microscopic description is in terms of finite-dimensional type I von Neumann algebra, for which there is no information loss. 

On the other hand, in the semi-classical description we have a good understanding of the in-falling experience. In particular, spacetime near the horizon is smooth so that any observer can simply fall through. A more algebraic perspective was recently provided by the work of Leutheusser and Liu \cite{Leutheusser:2021qhd, Leutheusser:2021frk}, who relate the existence of a half-sided modular inclusion \cite{Borchers:1991xk, Wiesbrock:1992mg} to an emergent time coordinate for the in-falling observer, see also \cite{deBoer:2022zps}. Notably, such half-sided modular inclusions only exist in type III von Neumann algebras \cite{Wiesbrock:1992mg}, so it seems that the emergence of time in the interior, and the existence of a sharp horizon, is closely related to the type III nature of the exterior algebra. Unfortunately, in this description the information recovery is not manifest: during the evaporation process information seems to get lost in the black hole interior. The tension between both descriptions, suited to either the type I or type III von Neumann algebra but not both, is at the heart of the black hole information problem.

These observations can be made particularly precise in the context of the AdS/CFT correspondence, where states associated to black holes in the bulk AdS space are related through the correspondence to states in a finite $c$ boundary CFT. Because the CFT is manifestly unitary, there is no information loss during the evaporation process. As pointed out above, this observation by itself does not resolve the information problem. The puzzle arises precisely when one tries to understand the information recovery process from a semi-classical bulk perspective in the limit $c\to \infty$. 

The observation by Witten \cite{Witten:2021unn} that the coupling to semi-classical gravity leads to a description in terms of a type II algebra provides a promising middle ground in which both the features of the emergent spacetime picture, as well as the microscopic quantum mechanical description can be reasonably well represented. The fact that the emergent operator algebra represents a type II von Neumann algebra makes it possible to study entanglement entropy in a well-defined way.\footnote{Type III von Neumann algebras do not have a trace or density matrices, so there is no meaningful definition of von Neumann entropy associated to a state.} Furthermore, we will argue that the type II description is well-suited to describe a black hole after the Page time. Our approach will be different from \cite{Witten:2021unn}, in the sense that the type II algebra does not enter through the crossed product construction. Instead, we will obtain an emergent type II$_1$ von Neumann algebra directly in the $G\to 0$ limit. See also the recent work \cite{Gomez:2024fij, Gomez:2024cjm} that discusses black hole evaporation in the context of type II$_1$ von Neumann algebras. 

In the $G\to 0$ limit, there is no gravitational back reaction, so it may seem that there is no content to the black hole information paradox. Indeed, by putting $G$ to zero a black hole with a given Schwarzschild radius formally carries infinite entropy and will never be able to fully evaporate. Nevertheless, as we will explain, there is still a well-posed version of the information paradox that remains to be resolved. Since in this limit one can still associate a non-zero Hawking temperature to the black hole, it remains possible to extract Hawking radiation from the black hole. This implies that, after a certain amount of radiation with energy $E$ has been collected, the entanglement entropy of the black hole has been reduced by a finite amount. In other words, the black hole has lost a certain amount of its quantum information.
Therefore, the $G\to 0$ version of the information paradox concerns the problem of explaining how the microscopic information contained in the black hole can be recovered from the radiation, while at the same time explaining why an in-falling observer still experiences a smooth horizon. We will argue that the framework of operator algebras is particularly well-suited to address this question.  

\subsection{Summary of results}

In this paper, we will formulate a version of the information puzzle and the corresponding recovery process in terms of type II$_1$ algebras. There is a direct relation to the Hayden-Preskill protocol as described by Yoshida and Kitaev \cite{Yoshida:2017non}, that we will carefully explain. We will make use of operator algebraic concepts that have been developed for type II algebras, in particular the index theory of Vaughan Jones \cite{Jones:1983kv}. The central idea will be to describe the evaporation process in terms of the (sub-)algebras associated to the black hole and the radiation and study how information is transferred from the black hole to the radiation algebra. We will find that the recovery protocol amounts to an implementation of Jones's basic construction \cite{Jones:1983kv, Longo:1989tt} and an operation called the canonical shift \cite{ocneanu1988quantized}. As we will explain, the canonical shift consists of the subsequent application of two modular conjugations, each associated to a different (sub-)algebra. At the microscopic level, it will become clear that this procedure involves an entanglement swap, that is implemented by so-called Jones projections. In this way, the recovery protocol can be interpreted as an algebraic quantum teleportation scheme; see, for example, \cite{huang2019dense, Conlon:2022slq} where quantum teleportation in the context of operator algebras and its relation to the canonical shift is discussed. Hence, on a microscopic level the information escapes from the black hole via a quantum teleportation step \cite{Verlinde:2013uja, Susskind:2017nto}.

A central ingredient in our description is the concept of a conditional expectation, which one can think of as an operator algebraic analogue of the partial trace that allows one to remove information from the algebra. Although the ingredients of the recovery protocol play an important role in the subfactor theory of type II$_1$ algebras, the notions of modular conjugation, conditional expectation and Jones index all have analogues in the type I and type III von Neumann algebra.\footnote{In particular, the index theory of Jones has been generalized to type III von Neumann algebras by Kosaki \cite{kosaki1986extension, kosaki1998type}, although the existence of conditional expectations (of finite index) for a given inclusion is by no means guaranteed.} For this reason, we expect the operator algebraic way of phrasing the puzzle, and its solution, to be useful in understanding the information recovery from a spacetime perspective.

The motivation and ultimate goal of this work is to connect the manifestly unitary microscopic description in terms of type I algebras and the emergent type III description in spacetime. The hope is that this will lead to a microscopic explanation of the QES prescription and the appearance of entanglement islands, and more generally reveal the mechanisms that are responsible for the resolution of the black hole information paradox within the spacetime description. In the present paper, we will not be able to complete this entire program, and will mostly focus on the description in terms of type I and type II algebras. However, we will point out an interesting connection between our recovery protocol and the recent discussion on the emergence of spacetime translations in the theory of half-sided modular inclusions for type III algebras. 

The outline of the paper is as follows. We begin in section \ref{sec:2} by describing the $G\to 0$ limit of the information paradox in operator algebraic language, and give a first description of the recovery protocol and its connection to the notion of the canonical shift. A more detailed description of the recovery protocol for type II algebras is presented in section \ref{sec:3} where we will make use of the mathematical theory of the basic construction and Jones projections. In section \ref{sec:4}, we will present a microscopic model of the information recovery protocol in terms of type I von Neumann algebras and make contact with the work of \cite{Yoshida:2017non}. The spacetime interpretation of the canonical shift in the context of half-sided modular inclusions is discussed in section \ref{sec:5}. We close with some concluding remarks in section \ref{sec:6}, and an appendix on the Jones index.

\section{Black Hole Information  and Operator Algebras}

\label{sec:2}

In this paper, we are interested in the microscopic process that is responsible for the information retrieval from an old black hole that is past the Page time. We will study this problem in the $G\to 0$ limit in which we simultaneously take the black hole mass $M$ to infinity so that the Schwarzschild radius and inverse Hawking temperature $\beta$ remain fixed.  For concreteness, we will put our discussion in the framework of AdS/CFT, but by formulating our arguments in the general language of operator algebras we hope to arrive at conclusions that are also valid for more general spacetimes.

\subsection{Black hole information in the $G\to 0$ limit}

The AdS/CFT setting we have in mind is as follows: we consider a generic state with high conformal dimension $M$ at very large $c$ with fixed ratio $M/c >\!\!> 1$. We identify the conformal dimension $M$ with the mass measured in terms of the AdS-scale. In the bulk this represents a semi-classical black hole whose radius is much larger than the AdS-scale. Since we are interested in the information content of the black hole it is appropriate to restrict the CFT Hilbert space to states with conformal dimension below a given value $M$. In that case, we are dealing with a finite-dimensional Hilbert space whose dimension can be expressed in terms of the entropy of the black hole
\begin{equation}
	\label{dimH}
	{\rm dim}\, {\cal H}_{M} = e^{S(M)}~.
\end{equation}

To model a black hole after the Page time we consider a CFT that is almost maximally entangled with a radiation bath. After tracing over the radiation Hilbert space one ends up with an almost maximally mixed state for the black hole with an invertible density matrix, which to a good approximation can be represented by\footnote{
The operators of the CFT will be identified with the algebra ${\cal B}({\cal H}_M)$ of bounded operators on the Hilbert space ${\cal H}_M$, and hence we will not be using any specific properties of CFT operators. We make the further simplifying assumption that all operator algebras correspond to factors and that subalgebras may also be identified with subfactors.}
\beq \label{eq:rhoBH}
\rho_{\rm BH} \,\approx\, \frac{\mathds{1}~}{\mathrm{dim}{\cal H}_M}~.
\eeq 
At this point we have entered the set-up considered by Hayden and Preskill. As was shown by these authors in \cite{Hayden:2007cs}, by collecting a certain amount of late radiation an observer, who is in possession of the entangled early radiation, can decode a part of the information that has fallen into the black hole at a much earlier time. The only requirement for the Hayden-Preskill protocol to work is that the in-falling information is sufficiently scrambled with respect to the late radiation. The question we will address is how this version of the information paradox can be formulated and possibly resolved in the language of operator algebras. 

One of the main points of the present paper is that the Hayden-Preskill version of the information paradox still persists in the $G\to 0$ limit, or equivalently in the $c\to\infty$ limit from the CFT perspective. In this limit, the dimension of the Hilbert ${\cal H}_{M}$ will become infinite. Furthermore, the density of states diverges, and for this reason there is no sensible definition of single pure black hole quantum state. But since the Hawking temperature stays fixed and finite, we can still consider the ratio of the dimensions of the Hilbert space for two different values of the mass $M$. In particular, if we consider a black hole of mass $M$ and a somewhat smaller black hole of mass $M-E$, the ratio of the dimensions of ${\cal H}_{M-E}$ and ${\cal H}_M$ is given by 
\begin{equation}
\label{boltzman}
		\frac{{\rm dim}\,{\cal H}_{M-E}}{{\rm dim}\,{\cal H}_{M}} = e^{S(M-E)-S(M)}= e^{-\beta E}~.
\end{equation}
While both Hilbert spaces become infinite-dimensional in the $G\to 0$ limit the relative size of the spaces is kept finite. As we will explain, in the language of von Neumann algebras the above quantity will be related to the index associated with the operator algebras of the larger and smaller black hole respectively. 

The interior degrees of freedom of the AdS black hole are described by a von Neumann algebra $\mathcal{M}$, which we identify with the operator algebra of the CFT restricted to the Hilbert space ${\cal H}_M$.  At finite $c$ and $M$ this algebra represents a type I von Neumann algebra, but in the $c\to \infty$ limit $\cal M$ will become a type II von Neumann algebra. In fact, since we restrict our Hilbert space to states below a given mass $M$ the operator algebra $\cal M$ associated to the black hole degrees of freedom corresponds to a type II$_1$ von Neuman algebra. To see this, note that by appropriately rescaling the trace by the dimension of the Hilbert space one obtains a normalized trace which obeys ${\rm tr}(\mathds{1})=1$. This is one of the defining properties of type II$_1$ algebra. 

To describe the structure of the emergent type II$_1$ von Neumann algebra in the $c\to \infty$ limit it is useful to purify the density matrix of the black hole \eqref{eq:rhoBH} by making use of the radiation degrees of freedom. Formally, this is achieved by applying the Gelfand-Naimark-Segal (GNS) construction to the algebra, which produces a Hilbert space ${\cal H}^{\mbox{\tiny \it GNS}}_{\cal M}$ whose states are in one-to-one correspondence with the operators in the algebra $\cal M$. This is most easily understood by going back to the finite $c$ situation: In this case it is clear that the dimension of the GNS Hilbert space is equal to the square of dimension of the corresponding CFT Hibert spaces
\beq
  \label{GNSsquare}
 {\rm dim}\, {\cal H}^{\mbox{\tiny \it GNS}}_{\cal M} = \bigl({\rm dim}\,{\cal H}_M\bigr)^2~.
\eeq 
One can think about ${\cal H}^{\mbox{\tiny \it GNS}}_{\cal M}$ as the space of all possible Bell states that can be made from the tensor product of two CFT's. The state that purifies the CFT density matrix corresponds to the identity operator, and can be represented as a ``generalized thermofield double state''\footnote{By this we mean a state of the form $\sum_n c_n|n\rangle|n\rangle$ with coefficients $p_n\equiv|c_n|^2$. For $p_n=e^{-\beta E_n}/Z$ it is the usual thermofield double state.}. The second CFT that purifies the original CFT should be thought of as being embedded in some complicated way inside the radiation Hilbert space. The GNS Hilbert space thus contains both the black hole degrees of freedom as well as the part of the old radiation that is entangled with it.
 
The size of the algebra $\cal M$, and hence the dimension of the GNS Hilbert space depends on the value of the mass $M$. We denote by $\cal N$ and ${\cal H}^{\mbox{\tiny \it GNS}}_{\cal N}$ the algebra and GNS Hilbert space associated to the smaller black hole with mass $M-E$. Since the Hilbert space ${\cal H}_{M-E}$ is contained in ${\cal H}_{M}$, we may regard $\cal N$ as a subalgebra of $\cal M$
\beq
\label{eq:inclusion1}
{\cal N}\subset {\cal M}~.
\eeq 
In the following, we will argue that the algebra $\cal N$ associated to the smaller black hole is not just a subalgebra of $\cal M$, but can be identified with a subfactor. This allows us to make use of Jones's index theory of type II$_1$ subfactors \cite{Jones:1983kv}. The index, which is denoted by $\left\lbrack {\cal M}:{\cal N}\right \rbrack$, is a measure of the relative size
of the algebras $\cal M$ and $\cal N$. It counts the number of linearly independent operators that need to be added to $\cal N$ to obtain the algebra $\cal M$.\footnote{The algebra $\cm$ is a finitely generated module over $\cn$ provided that the index is finite, and the corresponding orthonormal basis with dimension equal to the integer part of the index is the so-called Pimsner-Popa basis \cite{Pimsner1986}.} Heuristically, the Jones index can be thought of as the ratio of the dimensions of the corresponding GNS Hilbert spaces  
 \begin{equation}
	\left\lbrack {\cal M}:{\cal N}\right \rbrack= {{\rm dim}\, {\cal H}^{\mbox{\tiny \it GNS}}_{\cal M}  \over {\rm dim}\, {\cal H}^{\mbox{\tiny \it GNS}}_{\cal N} }~.
\end{equation}
Strictly speaking, this definition only applies to the case when the algebras in question are finite-dimensional matrix algebras. In the $c\to \infty$ limit, both dimensions are infinite, but their ratio is still well-defined. For a more precise definition of the Jones index we refer the reader to Appendix \ref{app:A} or some pedagogical references, see for example \cite{kosaki1991index,Jones:2015kbb,anantharaman2017introduction,banica2023principlesoperatoralgebras}.

In the present context, the index $\left\lbrack {\cal M}:{\cal N}\right \rbrack$ may be interpreted as a measure of the amount of information that can be extracted after an evaporation step from a black hole of mass $M$ to $M-E$ when an amount of energy $E$ has been released. By using the analogous relation to \eqref{GNSsquare} for the algebra $\cal N$, namely  ${\rm dim} {\cal H}^{\mbox{\tiny \it GNS}}_{\cal N} = \bigl({\rm dim}{\cal H}_{M-E}\bigr)^2$, and by using \eqref{boltzman} one finds that the index may be expressed in terms of the inverse Hawking temperature  $\beta$ and the energy $E$ via
\beq 
\left\lbrack {\cal M}:{\cal N}\right \rbrack= \exp{2\beta E}~.
\eeq

The fact that we can associate a finite index to an inclusion of infinite-dimensional algebras is a central property of type II$_1$ von Neumann algebras, and will be an essential ingredient in our discussion of the black hole information paradox and the recovery protocol. We will find that the Hayden-Preskill protocol has a natural formulation in terms of type II$_1$ operator algebras using Jones's basic construction and a procedure called the canonical shift. Before presenting the details of these constructions, we will first introduce the basic idea at a more heuristic level. 
    
\subsection{The algebras of Alice and Bob: the diary and  Hawking pairs}
  
The operator algebraic version of the information paradox in the $G\to 0$ limit can be stated as follows. We imagine that an early observer, called Alice, has hidden some information inside the black hole, which, following Hayden and Preskill, we will refer to as Alice's diary. The diary is small enough, so that its information content may be recovered after a relatively small amount of radiation with energy $E$ has been collected by another observer named Bob. As before, we denote the operator algebra of the black hole with mass $M-E$ after the evaporation step by $\mathcal N$, and will view it as a subalgebra of the algebra $\cal M$. The fact that the information is contained in $\cal M$, and not in $\cal N$, is formalized by saying that the information is contained in the relative commutant $\cal M\cap \cal N'$.\footnote{There is a slight subtlety in the sense that the size of the relative commutant and the index are related, but need not be equal (see, e.g., \cite{bakshi2023pimsner}). For now we will ignore this subtlety and assume that we are in a situation where both coincide.} Here, the commutant $\mathcal{N}'$ is defined as the part of the total algebra acting on the Hilbert space $\mathcal{H}^{\mbox{\tiny \it GNS}}_{\cal M} $ that commutes with $\cal N$. To make clear that this relative commutant contains Alice's diary we will denote it by
\beq
\label{eq:alicealgebra}
\mathcal{A} \equiv \mathcal{N}'\cap \mathcal{M}~.
\eeq
We think about the algebra $\mathcal{M}$ as acting on the combined Hilbert space of Alice's diary and a somewhat smaller black hole before Alice threw her diary in. Having access to the relative commutant $\mathcal{A}$ is the algebraic equivalent of having access to the diary. 

Let us now introduce Bob. Bob is in possession of the early radiation and, at first, has no access to the operators in the algebra $\cal M$, which contains  $\mathcal{A}$. In fact, the operators in Bob's possession all belong to the commutant $\cal M'$. The problem that Bob is faced with is that the radiation that he will receive after an evaporation step is not directly related to the operators contained in $\cal A$. Hence, he will not be able to simply read off the information contained in the diary. Furthermore, according to the semi-classical description, the process by which the Hawking radiation appears from the near-horizon region of the black hole amounts to pair creation of virtual particles that are mutually entangled, which means that they do not directly contain information about the black hole state. 
  
According to Hawking the algebra of the black hole gets enlarged, since after the evaporation step it also contains the partners of the emitted radiation. This means that instead of reducing the algebra from $\cal M$ to a smaller algebra,  we now obtain a larger algebra ${\cal M}_1$ that contains $\cal M$ as a subalgebra. In fact, we even obtain a sequence of inclusions of algebras 
  \beq \label{eq:inclusions}
\mathcal{M}\subset \mathcal{M}_{1}\subset \mathcal{M}_2~,
\eeq
where the algebra ${\cal M}_2$ is defined by also including the emitted Hawking radiation itself. Hence, both the in-falling as well as the outgoing particles of the Hawking pairs are contained in the largest algebra.  
We are now ready to define the operator algebra $\cal B$ that contains the emitted radiation with the help of which Bob plans to recover the information in the diary. It is given by the relative commutant 
 \beq
 \label{eq:bobalgebra}
 \mathcal{B}\equiv \mathcal{M}_1'\cap \mathcal{M}_2~.
 \eeq 
It will also be useful to introduce the algebra associated with the in-falling partner of Hawking pair, which will be denoted by
\beq
\label{eq:hawkingpartner}
\widetilde{\mathcal{B}}
\equiv \mathcal{M}'\cap \mathcal{M}_1~.
\eeq 

Comparing the earlier equations \eqref{eq:inclusion1} and \eqref{eq:alicealgebra}
to the equations \eqref{eq:inclusions}, \eqref{eq:bobalgebra} and \eqref{eq:hawkingpartner} presents us with a simplified version of the black information problem.  Indeed, the time evolution in \eqref{eq:inclusions} seems to go from left to right: the Hawking process creates at each time step a new particle pair that, when computed in effective field theory, leads to a larger algebra. However, since we are after Page time, unitarity requires that the actual internal algebra of the black hole becomes smaller, and hence the time evolution should actually go in the opposite direction. How do we reconcile these two apparently contrasting statements? 
 
Our goal will be to explain how the information in Alice's diary represented by the algebra $\cal A$ can eventually be recovered from the radiation, and reappear in the algebra $\cal B$. One key observation is that the algebras ${\cal M}_1$ and ${\cal M}_2$ are linked by the fact that after the evaporation step they contain either the in-falling particle or both the in-falling and outgoing particles of a Hawking pair. As we will now explain, this means that the inclusions of algebras (\ref{eq:inclusions}) represents an example of Jones's basic construction \cite{Jones:1983kv, Longo:1989tt}. The original route towards the basic construction involves the introduction of projections. We postpone the discussion of these Jones projections to the next section and here present a different characterization of the basic construction in terms of modular conjugation operators. 

\subsection{Information recovery via modular conjugation}
  
Let us consider the GNS Hilbert space ${\cal H}^{\mbox{\tiny \it GNS}}_{{\cal M}_1}$ associated to the algebra ${\cal M}_1$. The commutant algebra ${\cal M}_1'$ is represented on the same Hilbert space; its action is related to that of ${\cal M}_1$ by the so-called modular conjugation operator $J_{{\cal M}_1}$: 
  \beq
    \label{eq:M1JM}
 {\cal M}'_1 =  J_{{\cal M}_1}{\cal M}_1 \,J_{{\cal M}_1}~.
  \eeq 
While we postpone a precise definition of the modular conjugation operator to the next section, one can think about $J_{{\cal M}_1}$, physically, as mapping operators that act on one side of a maximally entangled ``vacuum'' state onto so-called ``mirror operators'' that create the same state by acting on the other side. These mirror operators have appeared before in the context of black hole interior reconstruction  \cite{Papadodimas:2012aq, Papadodimas:2013jku,Papadodimas:2015jra} (see also \cite{deBoer:2018ibj, DeBoer:2019yoe}), and we will think about the modular conjugation in the same way. In other words, it exchanges the operators associated to the outgoing and in-falling Hawking particles. 
 
The relationship between the algebras $\cal M$, ${\cal M}_1$ and ${\cal M}_2$ may be now be expressed as follows. The algebras ${\cal M}'$ and ${\cal M}_2$ can both be represented on ${\cal H}^{\mbox{\tiny \it GNS}}_{{\cal M}_1}$ and are mapped onto each other by the modular conjugation $J_{{\cal M}_1}$. We have
\beq
\label{eq:M2JM}
 {\cal M}_2 =  J_{{\cal M}_1}{\cal M}' \,J_{{\cal M}_1}~.
\eeq 
By combining the equations (\ref{eq:M1JM}) and \eqref{eq:M2JM} it is easy to verify that the modular conjugate of 
Bob's algebra $\cal B$ in  \eqref{eq:bobalgebra} is given by the algebra $\widetilde{\cal B}$, 
hence
\beq
\label{BJBJ}
{\cal B} = J_{{\cal M}_1} \widetilde{\cal B}\, J_{{\cal M}_1} 
\eeq 
This relation was to be expected since the algebra \eqref{eq:hawkingpartner} was identified with the in-falling Hawking particles. 

The central idea behind the information recovery protocol is to create a new situation in which we can repeat the basic construction once more, but now for the sequence of inclusions ${\cal N}\subset {\cal M} \subset {\cal M}_1$.  By following identical arguments as those leading to \eqref{BJBJ} one would find that Alice's algebra $\cal A$ is mapped on the same relative commutant $\mathcal{M}'\cap \mathcal{M}_1$ that we identified with the algebra $\widetilde{\cal B}$ of the in-falling Hawking partners, but now via the 
 the modular conjugation operator $J_{\cal M}$ associated to the GNS Hilbert space ${\cal H}^{\mbox{\tiny \it GNS}}_{{\cal M}}$. In this situation, one thus finds that 
\beq
\label{BJAJ}
\widetilde{\cal B} = J_{{\cal M}} {\cal A}\, J_{{\cal M}}~.
\eeq
The reconstruction map is then obtained by combining the two modular conjugation operators $J_{{\cal M}_1}$ and $J_{\cal M}$ into one operation $\Gamma$ by setting
\beq \label{eq:canonicalshift}
\Gamma(a)\equiv J_{\mathcal{M}_1}J_{\mathcal{M}}aJ_{\mathcal{M}}J_{\mathcal{M}_1}~, \qquad a\in \mathcal{A}~.
\eeq
As a consequence, Alice's algebra $\mathcal A$ is related to Bob's algebra $\mathcal B$ via 
\beq
{\cal B} =\Gamma(\cal A)~.
\eeq 
The operation in \eqref{eq:canonicalshift}, which plays an important role in the theory of type II von Neumann algebras, is called the canonical shift \cite{ocneanu1988quantized}, and leads to an isomorphism of algebras. 

As we will explain in detail below, the canonical shift is directly related to the implementation of the Hayden-Preskill protocol described by Kitaev and Yoshida \cite{Yoshida:2017non}, and may be interpreted as a quantum teleportation scheme \cite{huang2019dense,Conlon:2022slq}. A key ingredient in their construction is that Bob, by exploiting the operators associated to the late radiation can perform a projection on the combined state that also contains the early radiation. This projection leads to an entanglement swap and causes the algebra $\widetilde{\cal B}$, which was initially entangled with $\cal B$, to become entangled with the algebra $\cal A$ containing the information about the diary. The role of these projections will be further explained in the following section. 

We close this section with two comments. The first one relates to the question: Who or what determines the algebra $\cal A$ that can be decoded from the outgoing radiation? It is an important feature of the Hayden-Preskill protocol that Bob has a choice and can determine which part of the quantum information inside the black hole is recovered. In the present context, this means that the embedding ${\cal N}\subset {\cal M}$ of the algebra associated with the remaining black hole with mass $M-E$ is not uniquely fixed, but gets determined by a choice that can be made by Bob. How can this fact be understood in terms of the operator algebras? The decoding map $\Gamma$ in \eqref{eq:canonicalshift} consists of two parts. Let us first focus on the modular conjugation $J_{\mathcal{M}_1}$. Given the embedding ${\cal M}\subset {\cal M}_1$ it uniquely determines the embedding of the algebras ${\cal M}_1\subset {\cal M}_2$ via the basic construction. The modular conjugation $J_{\cal M}$, on the other hand, whose action determines the embedding of algebras ${\cal N}\subset {\cal M}$, is not uniquely determined by the embedding ${\cal M}\subset {\cal M}_1$. In the math literature, this step is called the downward basic construction \cite{Pimsner1986} and is defined up to a unitary transformation inside $\cal M$ (see also \cite{Pimsner1988, 10.1007/BF02392646}). To see this, consider the embedding ${\cal M}_1'\subset {\cal M}'$ viewed as algebras acting on the GNS Hilbert space ${\cal H}^{\mbox{\tiny \it GNS}}_{\cal M}$. This embedding can be modified by acting with an appropriate unitary $U'$ inside the algebra ${\cal M}'$, and hence this choice is controlled by Bob. Since $\cal N$ is obtained from ${\cal M'}$ via the modular conjugation,
\beq 
{\cal N} = J_{\cal M} {\cal M}_1' J_{\cal M}~,
\eeq 
we arrive at the conclusion that Bob can indeed determine which part of the algebra $\cal M$ is recovered. This point will become more clear in section \ref{sec:4}, when we discuss the connection with the Hayden-Preskill protocol. 

Our second comment concerns a possible spacetime interpretation. The notions of modular conjugation and sub-algebras apply to all types of von Neumann algebras, in particular to the emergent type III algebra that is discussed by Leutheusser and Liu. In the latter context, modular conjugations play a central role, since for a given half-sided modular inclusion ${\cal M}\subset {\cal M}_1$ they define a positive operator $P$ via the relation 
\beq 
J_{\mathcal{M}_1}J_{\mathcal{M}} = e^{-2iP}~.
\eeq  
The operator $P$ generates a family of unitaries $U(t)=e^{itP}$ for $t\in\mathbb{R}$, which in the physical setting of the eternal AdS black hole correspond to null translations when restricted to the horizon. The similarity between the definition of the canonical shift $\Gamma$ and the translation operator $e^{-2iP}$ suggests that the recovery map is also associated with a translation in spacetime. We will discuss this interpretation in more detail in section \ref{sec:5}. 

\section{Information Recovery and Jones's Basic Construction}

\label{sec:3}

In this section, we will discuss the recovery protocol in more detail. For this we will make use of Jones's basic construction formulated in terms of projection operators. In particular, we will show that Bob can recover the interior information when he is in possession of a suitable conditional expectation. This will allow him to perform the appropriate ``Bell measurement'' that leads to the entanglement swap required to decode the in-falling information from the outside Hawking radiation. In doing so, will make explicit use of the quantum state of the combined system of the black hole and the radiation, which is obtained by applying the GNS construction to the mixed state of the black hole. For finite $c$ the GNS Hilbert space simply consists of the tensor product of the black hole Hilbert space with that of the entangled radiation, and the purification of \eqref{eq:rhoBH} is simply given by the maximally entangled state. In the $c\to \infty$ limit, such a factorization does not exist, but the purification of the black hole density matrix does have a well-defined limit which is called the tracial state. It is the equivalent of the maximally entangled state for type II$_1$ algebras. 

Recall that states in the GNS Hilbert space are labeled by operators $a \in \mathcal{M}$. Given a separating\footnote{A state is called separating for $\mathcal{M}$ if the only operator in $\mathcal{M}$ which satisfies $a|\Psi\rangle=0$ is $a=0$.} state $|\Psi\rangle$ for the algebra $\mathcal{M}$, we can define states of the form 
\begin{equation} \label{eq:GNSstates}
|a\rangle \equiv a|\Psi\rangle~, \qquad a\in \mathcal{M}~.
\end{equation}
The inner product on such states is defined by the relation $
\langle b |a\rangle \equiv \langle \Psi|b^{\dagger}a|\Psi\rangle$. The GNS Hilbert space ${\cal H}
_{{\cal M}}$\footnote{Here and in the following, we will suppress the upper index and simply write $ {\cal H}_{\cal M}\equiv {\cal H}{} ^{\mbox{\tiny \it GNS}}_{\cal M}$.} is now obtained as the closure of the space of states \eqref{eq:GNSstates} with respect to the above inner product. The algebra $\mathcal{M}$ acts on the GNS Hilbert space via left multiplication
\beq 
a|b\rangle = |ab\rangle~.
\eeq
In general, the action of the algebra $\mathcal{M}$ on ${\cal H}_{{\cal M}}$ is not irreducible, which means that there exists a non-trivial commutant 
\beq 
\mathcal{M}'\equiv \{a' \in \mathcal{B}(\mathcal{H}_{\mathcal{M}})\,|\, [a',a]=0 \hspace{5pt} \mathrm{for \, all} \hspace{5pt} a\in \mathcal{M}\}~,
\eeq
which also acts on the same Hilbert space. Its action is defined via right-multiplication
\beq 
a'|b\rangle = |b (a')^{\dagger}\rangle~,
\eeq
where the Hermitian conjugate is there to make sure that the assignment defines a consistent action. 

As mentioned in the previous section, one can obtain the commutant from the original algebra via an anti-unitary operation $J_{\cal M}$ called modular conjugation. To define it, we first introduce the Tomita operator $S_{\cal M}:{\cal H}_{{\cal M}}\to{\cal H}_{{\cal M}}$ as the anti-linear map that implements the action of taking the Hermitian conjugate on the Hilbert space. In formulas, we have
\beq \label{eq:SM}
S_{\cal M} a|\Psi\rangle = a^{\dagger} |\Psi\rangle~.
\eeq
One can prove that the assignment \eqref{eq:SM} extends to a well-defined map on the full Hilbert space, with a unique polar decomposition of the form
\beq 
S_{\cal M} =J_{\cal M} \Delta_{\cal M}^{1/2}~.
\eeq
The positive-definite operator $\Delta_{\cal M}$ is called the modular operator, and gives rise to an important automorphism $a \mapsto \Delta^{-is}_{\cal M}a\Delta^{is}_{\cal M}$ of the algebra, the so-called modular flow that maps the algebra $\mathcal{M}$ to itself. The map $J_{\cal M}$ is the anti-unitary modular conjugation. From the definition, it immediately follows that 
\beq \label{eq:propJM}
J_{\mathcal{M}}^2 =1~, \qquad J_{\mathcal{M}}^{\dagger}=J_{\mathcal{M}}~.
\eeq
Since we are dealing with several algebras, we have assigned a sublabel $\cal M$ to the above maps to indicate that they are associated to the GNS Hilbert space of the algebra $\cal M$. The dependence on the state $|\Psi\rangle$ is left implicit. A main theorem of the Tomita-Takesaki theory \cite{Takesaki:1970aki} (we refer to  \cite{takesaki2003theory} for more details) now states that the commutant $\mathcal{M}'$ can be obtained from the original algebra $\mathcal{M}$ by conjugating with the modular conjugation map. In our physical set-up, the commutant algebra $\mathcal{M}'$ describes that part of the old Hawking radiation that is entangled with the black hole. As a consequence, the mirror operator $a' \in \mathcal{M}'$ associated to the operator $a\in \mathcal{M}$ can be obtained through the relation 
\beq 
a'= J_{\cal M} a J_{\cal M}~.
\eeq
The modular conjugation $J_{\cal M}$ is designed to find precisely these entangled degrees of freedom. 

In the present work, we will be mostly working in a type II$_1$ (or type I) algebra, where we may identify $|\Psi\rangle$ with the tracial (or maximally entangled) state, which satisfies
\begin{equation} 
	\langle \Psi | a |\Psi\rangle = \mathrm{tr}_{\mathcal{M}}(a)~,
\end{equation}
where $\mathrm{tr}_{\mathcal{M}}$ is the unique normalized trace on $\mathcal{M}$ which satisfies $\tr_{\mathcal{M}}(\mathds{1})=1$. The existence of such a trace is guaranteed for type II$_1$ algebras. In this situation, the modular operator is equal to the identity, $\Delta \equiv  {\mathds 1}$, (as can be seen, for example, from the KMS condition) and hence the modular flow is trivial. Therefore, the modular conjugation operator itself is simply given by  
\begin{equation} 
J_{\mathcal{M}}:\mathcal{H}_{\mathcal{M}}\to \mathcal{H}_{\mathcal{M}}~: \hspace{10pt} |a\rangle \mapsto |a^\dagger\rangle~.
\end{equation}

Let us return to the inclusion of algebras \eqref{eq:inclusions} that was obtained from $\cal M$ by adding the in-falling and emitted Hawking partners to the algebra. Clearly, the GNS Hilbert space ${\cal H}_{\cal M}$ is contained as a linear subspace inside the larger GNS Hilbert space ${\cal H}_{{\cal M}_1}$. This implies that there exists a projector $e_{\cal M}$ that maps ${\cal H}_{{\cal M}_1}$ onto the subspace ${\cal H}_{\cal M}$ 
\beq
e_{\cal M}: {\cal H}_{{\cal M}_1} \to {\cal H}_{\cal M}~.
\eeq 
The operator $e_{\cal M}$ is called the Jones projection. The essential idea of Jones theory is that the projector $e_{\cal M}$ commutes with the algebra $\cal M$, but is not a part of the algebra ${\cal M}_1$. Hence, by including the projector $e_{\cal M}$ one obtains a larger algebra that contains both $\cal M$ and ${\cal M}_1$. This extended algebra turns out to be exactly given by the algebra ${\cal M}_2$ that contains the outgoing Hawking partner. Physically, the Jones projection $e_{\cal M}$ corresponds to putting the Hawking pairs back into their entangled ground state. In other words, it is a Bell projection onto the vacuum state near the black hole horizon.  

From a microscopic perspective the Hawking process is unitary in the sense that the actual algebra of the black hole, which has lost a bit of its mass in the evaporation step, should have become smaller. As before, we denote the relevant algebra by $\mathcal{N}$ and its GNS representation by $\mathcal{H}_{\mathcal{N}}$. In order for the information in the relative commutant ${\cal A}={\cal N}'\cap {\cal M}$ to be recoverable from the radiation, one has to arrange things so that the algebra ${\cal M}_1$ is obtained by applying the basic construction to the embedding ${\cal N}\subset {\cal M}$. This implies that ${\cal M}_1$ must contain a projector $e_{\mathcal{N}}$ that takes the GNS Hilbert space $\mathcal{H}_{\mathcal{M}}$ and projects it onto the smaller subspace $\mathcal{H}_{\mathcal{N}}$. In the next subsection, we will describe how the information of the diary that is contained in the algebra $\cal A$ can be retrieved from the outgoing radiation using the Jones projections. We will again find that the embedding $\mathcal{N}\subset {\cal M}$, and hence which part of the black hole information is retrieved, is not uniquely fixed by the embedding ${\cal M}\subset {\mathcal{M}}_1$: It is determined up to a unitary that can be controlled by Bob.

\subsection{Conditional expectations and Jones's basic construction}

\newcommand{\vxi}{\xi}

Since Bob is located outside of the black hole horizon, he does not have access to the operators $\mathcal{M}$ that involve the interior degrees of freedom. In particular, he cannot use the operators in $\cal A$ to access the diary state. We assume, however, that Bob has managed to capture the Hawking radiation during the evaporation process, and that he can act with arbitrary operators on this radiation using some powerful quantum computer. In other words, we assume that he has access to the commutant algebra $\mathcal{M}'$. What are the additional ingredients that Bob needs to reconstruct the information of the diary? 

We will now show that for Bob to be able to recover the information contained in the algebra $\cal A$, he needs to be able to apply a conditional expectation
	\begin{equation} \label{eq:conditionalexpectation}
	\mathcal{E}:\mathcal{M}\to \mathcal{N}~,
	\end{equation} 
that effectively projects out the information of the diary from the interior of the black hole. This allows him to carry out the algebraic equivalent of a quantum teleportation protocol that transfers the information to the radiation degrees of freedom. 

One should think about the conditional expectation $\cal E$ as the algebraic equivalent of taking a partial trace over the degrees of freedom associated to the diary. An important property of the conditional expectation is that it satisfies 
\beq \label{eq:bimoduleproperty}
\mathcal{E}(b_1ab_2)=b_1\mathcal{E}(a)b_2 \qquad \mathrm{for} \qquad a\in \mathcal{M}, \, b_1,b_2 \in \mathcal{N}~,
\eeq 
which is called the bimodule property. Furthermore, it is compatible with the Hermitian conjugate in the sense that $\mathcal{E}(a^{\dagger})=\mathcal{E}(a)^{\dagger}$, and it is normalized such that $\mathcal{E}(\mathds{1})=\mathds{1}$. The conditional expectation \eqref{eq:conditionalexpectation} can now be used to define the Jones projection $e_{\cal N}$ via the assignment
\begin{equation} \label{eq:definitionprojector}
e_{\mathcal{N}}:\mathcal{H}_{\mathcal{M}} \to \mathcal{H}_{\mathcal{N}}~, \hspace{10pt} e_{\mathcal{N}}|a\rangle  = |\mathcal{E}(a)\rangle~.
\end{equation}
From the above properties of the conditional expectation it is easy to derive that $e_{\mathcal{N}}^2 =e_{\mathcal{N}} = e_{\mathcal{N}}^{\dagger}$, so that $e_{\mathcal{N}}$ indeed defines a projector, which removes the diary Hilbert space from $\mathcal{H}_{\mathcal{M}}$. 

For Bob to be able to apply the projector, he first needs to extend his algebra of observables. Because $e_{\mathcal{N}}$ is \emph{not} part of $\mathcal{M}$, we define a somewhat larger algebra $\mathcal{M}_1$ that is obtained from the algebra $\cal M$ by also including the projector $e_{\cal N}$. To be precise, we set 
\begin{equation} \label{eq:M1projector}
\mathcal{M}_{1}\equiv \langle \mathcal{M}, e_{\mathcal{N}} \rangle~,
\end{equation}
where the brackets indicate that we consider the smallest von Neumann algebra that contains both $\mathcal{M}$ and $e_{\mathcal{N}}$. This can be achieved by taking the double commutant of the union, $(\mathcal{M}\cup \{e_{\mathcal{N}}\})''$, within the Hilbert space $\mathcal{H}_{\mathcal{M}}$, which is automatically a von Neumann algebra. As a result, we obtain the following sequence of inclusions
\begin{equation} 
	\mathcal{N} \subset \mathcal{M} \subset_{e_{\mathcal{N}}} \mathcal{M}_1~,
\end{equation}
where we have used the symbol $\subset_{e_{\mathcal{N}}}$ to stress that $\mathcal{M}_1$ is obtained from $\mathcal{M}$ by including the projector $e_{\mathcal{N}}$. The algebra $\mathcal{M}_1$ is known as the basic extension \cite{Jones:1983kv}, and was first introduced by Jones to study subfactors of type II$_1$ algebras. An important property of the above construction is that when $\mathcal{N}\subset \mathcal{M}$ is an inclusion of type II$_1$ factors the basic extension is automatically a type II$_1$ factor as well. Furthermore, the Jones index associated to the inclusion is preserved,
\begin{equation} \label{eq:indexpreserved}
	[\mathcal{M}:\mathcal{N}]=[\mathcal{M}_1:\mathcal{M}]~.
\end{equation}
The proof of these statements can be found in
\cite{Jones:1983kv} (see also Appendix \ref{app:A} or \cite{Pimsner1986}). As a consequence, one can repeat the basic construction for the inclusion $\mathcal{M}\subset \mathcal{M}_1$ to obtain an even larger algebra, that we denote by $\mathcal{M}_2$. Continuing in this fashion, one constructs an infinite sequence of algebra inclusions that is called the Jones tower. 

The above procedure for obtaining the algebra $\mathcal{M}_1$ in terms of the Jones projector is equivalent to the one we presented in section \ref{sec:2} in terms of the commutant of $\mathcal{N}$ within the GNS Hilbert space of $\mathcal{M}$, and the modular conjugation operator $J_{\mathcal{M}}$. To be precise, we can express the algebra in \eqref{eq:M1projector} also via the formula 
\begin{equation} \label{eq:M1}
\mathcal{M}_1=J_{\mathcal{M}}
\mathcal{N}'J_\mathcal{M}~.
\end{equation}
Let us check that this is indeed the case. We first introduce a useful characterization for being an element of the subalgebra $\mathcal{N}$. One can show that for $b\in \mathcal{M}$ it holds that
\beq \label{eq:charN}
b\in \mathcal{N} \qquad \longleftrightarrow \qquad  e_{\mathcal{N}}b=be_{\mathcal{N}}~.
\eeq
To show this, one has to prove both implications. Assume first that $b\in \mathcal{N}$. Then, it follows for $a\in \mathcal{M}$ that 
\begin{equation} 
e_{\mathcal{N}}b|a\rangle =\mathcal{E}(ba)|\Psi\rangle = b \mathcal{E}(a)|\Psi\rangle = b e_{\mathcal{N}}|a\rangle~,
\end{equation}
where we have used the definition of the projector in terms of the conditional expectation \eqref{eq:definitionprojector} and its bimodule property \eqref{eq:bimoduleproperty}. Because the GNS Hilbert space is generated by states of the form $|a\rangle$ with $a\in \mathcal{M}$, this proves that $e_{\mathcal{N}}b=be_{\mathcal{N}}$ as required. Conversely, if $e_{\mathcal{N}}b=be_{\mathcal{N}}$ it follows that 
\begin{equation} 
\mathcal{E}(b)|\Psi\rangle = e_{\mathcal{N}}b|\Psi \rangle = be_{\mathcal{N}} |\Psi\rangle   = b|\Psi \rangle~.
\end{equation}
Using that the state $|\Psi\rangle$ is separating we conclude that $b\in \mathcal{N}$. From \eqref{eq:charN} it immediately follows that 
\begin{equation} \label{eq:B=}
\mathcal{N} = \{e_{\mathcal{N}}\}'\cap \mathcal{M} = (\{e_{\mathcal{N}}\}\cup \mathcal{M}')'~.
\end{equation}
By taking another commutant of the above expression we obtain the identity 
\begin{equation} \label{eq:identityN'}
\mathcal{N}'=\langle \mathcal{M}', e_{\mathcal{N}} \rangle~.
\end{equation}
Finally, to arrive at \eqref{eq:M1} one needs to conjugate both sides of equation \eqref{eq:identityN'} by $J_{\mathcal{M}}$. Clearly, the left-hand side becomes the algebra $J_{\mathcal{M}}\mathcal{N}'J_{\mathcal{M}}$. To simplify the right-hand side we use the identity $J_{\mathcal{M}}e_{\mathcal{N}}J_{\mathcal{M}}= e_{\mathcal{N}}$, which follows from 
\begin{equation} \label{eq:JeJ}
e_{\mathcal{N}}J_{\mathcal{M}} |a\rangle = \mathcal{E}(a^{\dagger})|\Psi\rangle =  \mathcal{E}(a)^{\dagger}|\Psi\rangle = J_{\mathcal{M}} e_{\mathcal{N}} |a\rangle~.
\end{equation}
Using this we conclude that $J_{\mathcal{M}}\langle \mathcal{M}', e_{\mathcal{N}} \rangle J_{\mathcal{M}}=\langle J_{\mathcal{M}}\mathcal{M}'J_{\mathcal{M}}, J_{\mathcal{M}}e_{\mathcal{N}}J_{\mathcal{M}} \rangle = \mathcal{M}_1$, which proves  \eqref{eq:M1}. At this point, we have gathered all the ingredients to discuss the recovery protocol for type II$_1$ von Neumann algebras in the language of Jones projections.

\subsection{Information recovery using the Jones projections}

Alice's action of creating her diary, and afterwards throwing it into the old black hole, can be modeled by acting with some operator $\vxi \in \cal A$ on the state $|\Psi\rangle$, where from now on we denote by $|\Psi\rangle$ the tracial state in the GNS Hilbert space $\mathcal{H}_{{\cal M}_1}$ of the algebra $\mathcal{M}_1$. We define the corresponding state of the black hole as
\begin{equation} 
|\vxi\rangle \equiv \,\vxi|\Psi \rangle~.
\end{equation}
Our second observer Bob wants to retrieve the information of the diary from the black hole. On the level of the algebras, this information is encoded in correlation functions of the form
\begin{equation} \label{eq:correlationAlice}
\langle a \rangle_\vxi \equiv \langle \vxi|a|\vxi\rangle ~, \qquad a\in \cal A~.
\end{equation} 
However, as mentioned before Bob does not have direct access to the algebra $\mathcal{A}$. Instead, we assume hat Bob can apply the projection $e_{\mathcal{N}}\in \mathcal{M}_1$ to obtain the state 
\begin{equation} 
\label{eq:projectedstate}
	e_{\mathcal{N}}|\vxi\rangle = |e_{\mathcal{N}}\vxi\rangle \in \mathcal{H}_{\mathcal{M}_1}~.
\end{equation}
This projector becomes available to Bob after an evaporation step. Note that we can write the projector $e_{\mathcal{N}}$ inside the ket-state, because we are working in the GNS Hilbert space of $\mathcal{M}_1$. It is important to notice that $e_{\mathcal{N}}$ has removed the diary information from all observers that only have access to $\mathcal{M}_1$, but the information is still contained in the state $|e_\mathcal{N}\vxi\rangle$ provided that one can act with an operator that changes $\vxi$ to $a\vxi$. To achieve this, Bob needs to transfer operators into the radiation algebra $\mathcal{B}$.

As was mentioned in section \ref{sec:2}, this information transfer goes in two steps and uses the intermediate channel $\widetilde{\mathcal{B}}$, that represents the in-falling Hawking particle. First, we show how to move operators from $\mathcal{A}=\mathcal{N}'\cap \mathcal{M}$ into $\widetilde{\mathcal{B}}=\mathcal{M}'\cap \mathcal{M}_1$. For $a\in \mathcal{A}$ we define the map
\begin{equation} 
\gamma_{\mathcal{M}}(a)\equiv J_{\mathcal{M}}a^{\dagger}J_{\mathcal{M}}~,
\end{equation}
which, by virtue of the expression \eqref{eq:M1} for the basic extension, takes values in $\widetilde{\mathcal{B}}$. Using the properties of the modular conjugation in \eqref{eq:propJM} it is now easy to verify that for $a,b\in \cM$ the map $\gamma_{\mathcal{M}}$ satisfies 
\begin{equation} \label{eq:antihom}
\gamma_{\mathcal{M}}(ab)=\gamma_{\mathcal{M}}(b)\gamma_{\mathcal{M}}(a)~, \qquad  \gamma_{\mathcal{M}}(a^{\dagger})=\gamma_{\mathcal{M}}(a)^{\dagger}~.
\end{equation}
As a result, it defines an anti-homomorphism\footnote{An \emph{anti-}homomorphism is a map that reverses the order of the product in the algebra.} on the relative commutants
\begin{equation} 
\gamma_{\mathcal{M}}:\mathcal{A} \to \widetilde{\mathcal{B}}~.
\end{equation} 
To ensure that the information transfer works we need the algebras $\cal A$ and $\cal \widetilde{\mathcal{B}}$ to be ``entangled'' in a certain way. This entanglement is realized by the projection operator $e_{\cal N}$ which, in physics language, represents a maximally entangled Bell pair between the algebras $\mathcal{A}$ and $\widetilde{\mathcal{B}}$. Formally, this is expressed through the following two important identities
\begin{equation} \label{eq:gammaprojector}
a e_{\mathcal{N}}=\gamma_{\mathcal{M}}(a)e_{\mathcal{N}}~, \qquad e_{\mathcal{N}}a =e_{\mathcal{N}}\gamma_{\mathcal{M}}(a)~.
\end{equation}
The proof is straightforward. For $a \in \cal{A}$ and $b\in \mathcal{M}$ we indeed have 
\begin{equation} 
a e_{\mathcal{N}}|b\rangle = |a\mathcal{E}(b)\rangle = |\mathcal{E}(b)a\rangle = \gamma_{\mathcal{M}}(a)|\mathcal{E}(b)\rangle = \gamma_{\mathcal{M}}(a)e_{\mathcal{N}}|b\rangle~,
\end{equation}
which proves the first identity in \eqref{eq:gammaprojector}. A similar argument can be used to derive the second identity. The physical interpretation of \eqref{eq:gammaprojector} in terms of entanglement will become more clear in the next section, when we discuss the microscopic model.

The second step in the recovery protocol is simply a repetition of the first step, but now for the inclusion $\mathcal{M}\subset \mathcal{M}_1$, within the GNS Hilbert space of $\mathcal{M}_1$. For an operator $b\in \widetilde{\mathcal{B}}=\mathcal{M}'\cap \mathcal{M}_1$ we define the map
\begin{equation} 
\gamma_{\mathcal{M}_1}(b)\equiv J_{\mathcal{M}_1}b^{\dagger}J_{\mathcal{M}_1}~,
\end{equation}
where $J_{\mathcal{M}_1}$ is the modular conjugation defined with respect to the tracial state on the GNS Hilbert space $\mathcal{H}_{\mathcal{M}_1}$. Let us now check in which part of the algebra the information can be found. Via a similar argument as before, on can show that the map $\gamma_{\mathcal{M}_1}$ takes values in the commutant of $\mathcal{M}_1$. To be precise, one needs to introduce the corresponding Jones projection $e_{\cal M}: {\cal H}_{{\cal M}_1} \to {\cal H}_{\cal M}$, and construct the extension algebra 
\begin{equation} 
\mathcal{M}_2 \equiv \langle \mathcal{M}_1,e_{\mathcal{M}} \rangle~. 
\end{equation}
As was already pointed out in \eqref{eq:M2JM}, it now follows that $\gamma_{\mathcal{M}_1}$ maps the operators $\widetilde{\mathcal{B}}=\mathcal{M}_1\cap \mathcal{M}'$ acting on the in-falling Hawking particles to the operators $\mathcal{B}=\mathcal{M}_2\cap \mathcal{M}_1'$ that act on the outgoing ones. This is the algebra that Bob has has access to. Note that the map $\gamma_{\mathcal{M}_1}$ has similar properties as the map $\gamma_{\mathcal{M}}$ in \eqref{eq:antihom}.  

We can now combine both steps and arrive at the final result for our recovery map $\Gamma$ as the composition of $\gamma_{\mathcal{M}_1}$ and $\gamma_{\mathcal{M}}$,
\begin{equation} 
\Gamma: \mathcal{M}\cap \mathcal{N}' \overset{\gamma_{\mathcal{M}}}{\longrightarrow} \mathcal{M}_1\cap \mathcal{M}' \overset{\gamma_{\mathcal{M}_1}}{\longrightarrow} \mathcal{M}_2\cap \mathcal{M}_1'~.
\end{equation}
When we write this out in terms of the modular conjugation operators, it can be expressed explicitly as 
\begin{equation} 
\Gamma(a)  \equiv  \gamma_{\mathcal{M}_1}\circ \gamma_{\mathcal{M}}(a)= J_{\mathcal{M}_1}J_{\mathcal{M}}aJ_{\mathcal{M}}J_{\mathcal{M}_1}~.
\end{equation}
Hence, the map $\Gamma$ is precisely the canonical shift that we already  mentioned in the previous section. To show that it defines a recovery map, we have to argue that Bob can reproduce the correlation functions in \eqref{eq:correlationAlice} using operators in $\mathcal{B}$. This follows from \eqref{eq:gammaprojector}. In fact, the information recovery amounts to the following identity
\begin{equation} 
\label{eq:recovery1}
\Gamma(a)|e_\mathcal{N}\vxi\rangle = e_{\mathcal{N}}|\vxi\gamma_{\mathcal{M}}(a)\rangle = e_{\mathcal{N}}\gamma_{\mathcal{M}}(a)|\vxi\rangle = e_{\mathcal{N}}a|\vxi\rangle = |e_{\mathcal{N}}a\vxi\rangle~,
\end{equation}
where we have used the fact that $\gamma_{\mathcal{M}}(a)\in \mathcal{M}'$ commutes with $\xi\in \mathcal{A}$. As a consequence, after Bob has applied the the projector $e_{\mathcal{N}}$ to the state $|\xi\rangle$ of the black hole, he can then use the canonically shifted operator $\Gamma(a)$ to compute an expectation value of the form
\begin{equation} 
\label{eq:recovered}
\langle e_\mathcal{N}\vxi| \Gamma(a)|e_\mathcal{N}\vxi\rangle = \langle \vxi|e_{\mathcal{N}}a|\vxi\rangle = \mathrm{tr}_{\mathcal{M}_1}(e_{\mathcal{N}}a\vxi\vxi^{\dagger}) = [\mathcal{M}:\mathcal{N}]^{-1}\mathrm{tr}_{\mathcal{M}}(a\vxi\vxi^{\dagger})~,
\end{equation}
where at the last step we have used that
\beq \label{eq:traceM1}
\mathrm{tr}_{\mathcal{M}_1}(e_{\mathcal{N}}a)=[\mathcal{M}:\mathcal{N}]^{-1}\,\mathrm{tr}_{\mathcal{M}}(a)~,
\eeq
as follows from the uniqueness of the normalized trace on $\mathcal{M}$.

\begin{figure}
	\centering
	\includegraphics[width=0.40\linewidth]{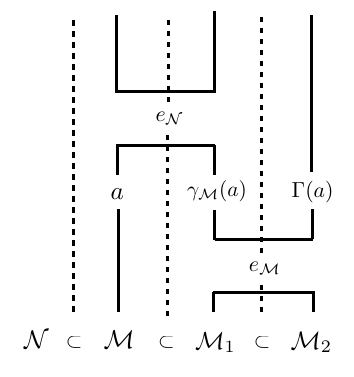}
	\caption{A pictorial representation of the information recovery protocol. We have indicated the algebras $\mathcal{N}\subset \mathcal{M}\subset \mathcal{M}_1 \subset \mathcal{M}_2$ obtained from the basic construction by striped lines. The canonical shift $\Gamma$ acts in two steps, first using the entanglement property \eqref{eq:gammaprojector} of the Jones projection $e_\mathcal{N}$, and afterwards the equivalent for $e_\mathcal{M}$. In this way, the canonical shift $\Gamma$ transfers operators from the Alice's algebra $\mathcal{A}=\mathcal{N}'\cap \mathcal{M}$ to Bob's algebra $\mathcal{B}=\mathcal{M}_1'\cap \mathcal{M}_2$ using the intermediate channel $\widetilde{\mathcal{B}}=\mathcal{M}'\cap \mathcal{M}_1$.} 
	\label{fig:definitiongamma}
\end{figure}

After properly normalizing the state $|e_\mathcal{N}\vxi\rangle$, Bob can therefore reproduce all correlation functions in \eqref{eq:correlationAlice} using operators $\Gamma(a)\in \mathcal{B}$. Moreover, it is clear from the properties of the maps $\gamma_{\mathcal{M}}$ and $\gamma_{\mathcal{M}_1}$ that the canonical shift preserves the algebraic relations between the operators 
\begin{equation}
\Gamma(ab)=\Gamma(a)\Gamma(b)~, \hspace{10pt} \Gamma(a^{\dagger})=\Gamma(a)^{\dagger}~.
\end{equation} 
In that sense, it recovers the information of the diary from the radiation. It is useful to have a pictorial representation of the map $\Gamma$ in mind, see Figure \ref{fig:definitiongamma}.

It is worth pointing out that the language that we use here is very similar to the one used in the approach of holographic quantum error correction in the context of infinite-dimensional von Neumann algebras as initiated in \cite{Kang:2018xqy} (see also \cite{Beny:2007ewj, Kang:2019dfi, Hollands:2020owv, Faulkner:2020iou, Furuya:2020tzv, Gesteau:2021jzp}). In particular, it was argued in \cite{Faulkner:2020hzi} that one can view the holographic map between boundary and bulk degrees of freedom as a conditional expectation. The Jones basic construction is also mentioned briefly in that context. It would be interesting to see if one can think about our recovery protocol more generally in terms of this algebraic version of the holographic map in the AdS/CFT correspondence.

\subsection{Scrambling and Bob's choice}

One important issue that we have not discussed yet is the role of scrambling. A related question is the following: How does Bob get access to the conditional expectation $e_{\cal N}$ that allows him to retrieve the information about Alice's diary? Or phrased in a different way: How does Bob control which part of the information contained in the black hole can be recovered from the radiation? Clearly, the information in the diary does not simply get emitted by the black hole, but at best is released in some scrambled form. 

Scrambling can be modeled by applying a random unitary $U$ on the state of the diary. This unitary $U$ is part of the algebra $\cal M$ and hence it changes the embedding ${\i: {\cal N} \hookrightarrow {\cal M}}$ of the algebra ${\cal N}$ as a subalgebra of ${\cal M}$. This also changes the algebra $\cal A$ by the same unitary conjugation: ${\cal A} \to U {\cal A} U^\dagger$. 
 Scrambling thus affects the state $|\vxi\rangle$ containing the diary and changes it into  
\beq
\label{eq:scrambled}
|\vxi\rangle\  \rightarrow \ U |\vxi\rangle \qquad \mbox{with} \qquad UU^\dagger ={\mathds 1}~, \quad U\in {\cal M}~.
\eeq 
In the following, we assume that Bob knows which unitary $U$ has acted on the state $|\vxi\rangle$. His problem is to find a way to undo the scrambling. 
It will be useful to represent the new scrambled state \eqref{eq:scrambled} in the GNS Hilbert space ${\cal H}_{{\cal M}_1}$ of the enlarged algebra ${\cal M}_1$ that contains the Jones projection $e_{\cal N}$.
If Bob would again apply this same projector and follow the identical steps, he would not obtain the correct correlation functions as in \eqref{eq:recovered}. Ideally, Bob would like to have access to a new projector $
e^{U}_{\cal N} \equiv U e_{\cal N}U^\dagger $,  
which is the Jones projection associated to the new embedding of the algebra $\cal N$. The problem is that Bob cannot make use of the entire algebra $\cal M$. Hence, he seems to be stuck with the same projector $e_{\cal N}$. 

But Bob can construct a new projector by using a conjugate unitary $U'$ that is part of ${\cal M}'$, 
\beq 
e^{U'}_{\cal N}\equiv \,U'{}^\dagger e_{\cal N}U'\qquad \mbox{with}\qquad U' = J_{\cal M} UJ_{\cal M}~, \quad U' \in {\cal M}'~.
\eeq 
By acting on the scrambled state \eqref{eq:scrambled} with the 
projector $e^{U'}_{\cal N}$ one finds that the original \eqref{eq:projectedstate} gets replaced by    
\beq
|e_{\cal N}\vxi\rangle \ \to \ e^{U'}_{\cal N} |U \vxi\rangle~. 
\eeq 
The resulting state appears to be quite different, but it turns out that the recovery protocol can still be made to work if $U$ is indeed scrambling and $\vxi$ is an operator of small rank compared to the square root of the index $[{\cal M}:{\cal N}]$. To show this let us go through the analogous steps as in \eqref{eq:recovery1}. One obtains the following expression
\begin{equation} 
\label{eq:recoveryscrambled}
\langle U\vxi| e^{U'}_{\cal N}\Gamma(a)e^{U'}_{\cal N} |U\vxi\rangle = \langle U\vxi|e^{U'}_{\cal N} |U\vxi\gamma_{\mathcal{M}}(a)\rangle= \langle \vxi | U^\dagger U'{}^\dagger e_{\cal N}  U' U \gamma_{\mathcal{M}}(a) |\vxi\rangle~.
\end{equation}
It is easy to see that the state $|\Psi\rangle$ obeys $U' U|\Psi \rangle = UU^\dagger|\Psi \rangle = |\Psi\rangle$. Using the fact that the right hand side also contains the projector $e_{\cal N}$ one can argue that the following identity holds to good approximation
\beq
\langle \vxi | U^\dagger U'{}^\dagger e_{\cal N}  U' U \gamma_{\mathcal{M}}(a) |\vxi\rangle \approx \langle \vxi|e_{\cal N} \gamma_{\mathcal{M}}(a)|\vxi \rangle~.
\eeq 
The rest of the recovery protocol then proceeds as before. In order to show that this approximation holds, one uses that the matrix $U$ is a scrambling unitary. The error in the last expression will be explained in a bit more detail in the following section, when we discuss the recovery protocol in the context of a microscopic model of type I von Neumann algebras. 

\section{A Type I Operator Algebraic Microscopic Model}
\label{sec:4}

To the uninitiated reader, our algebraic approach to black hole information in terms of type II$_1$ algebras might seem a bit abstract. To make the underlying physical ideas more manifest, it is useful to work things out explicitly in a putative microscopic theory described by a finite number of black hole microstates. In that case, the algebra of observables is a type I von Neumann algebra, and we will argue that the procedure described above, in terms of the Jones projectors and the canonical shift, can actually be thought of as an algebraic formulation of a quantum teleportation protocol. In this language, the projectors that are used in constructing the relevant algebras correspond to entangled Bell pairs that are shared between the inside and the outside of the black hole, and are essential in making information transfer possible. 

In this section, we will present such an explicit microscopic model in terms of Hilbert spaces and unitary gates that model the evaporation step. We will write down an (approximate) recovery map that can be used, after Page time, to decode the Hawking radiation, and we will find that it can be neatly expressed in terms of the canonical shift. In terms of the corresponding type I von Neumann algebras, the physical meaning of the recovery protocol will become clear: It amounts to a teleportation step in the spirit of the Hayden-Preskill protocol. In particular, this discussion illustrates the important role that entanglement and scrambling play in recovering the quantum information.

\subsection{A black hole after Page time}

Our set-up involves an old black hole after Page time. In the microscopic theory, the resulting state after creation of the black hole and its partial evaporation takes the form
\begin{equation}
|\chi\rangle =\frac{1}{\sqrt{d_M}} \sum_{i}|i\rangle |\chi_i\rangle~,
\end{equation}
where the black hole microstates are labeled by $|i\rangle \in \mathcal{H}_M$, and the states $|\chi_i\rangle$ are radiation states in an auxiliary Hilbert space $\mathcal{H}_{R}$. The number of states is denoted by 
\begin{equation} 
d_{M}\equiv \dim \mathcal{H}_M~.
\end{equation} 
The precise form of the radiation states $|\chi_i\rangle$ depends on the early history of the black hole and the microscopic dynamics of the evaporation process. The assumption that the black hole is old implies that the radiation Hilbert space $\mathcal{H}_R$ is much larger than the black hole Hilbert space $\mathcal{H}_M$,
\begin{equation} 
\dim \mathcal{H}_R \gg \dim \mathcal{H}_M~.
\end{equation}
As a consequence, the radiation states are approximately orthogonal
\begin{equation} \label{eq:orthogonal}
\langle \chi_i |\chi_j \rangle \approx \delta_{ij}~.
\end{equation} 
To compute the density matrix of the black hole we need to trace over the radiation system, and using the orthogonality relation \eqref{eq:orthogonal} the resulting density matrix is approximately proportional to the identity matrix
\begin{equation} \label{eq:oldBH}
\rho_{\rm BH}
\approx \frac{1}{d_M} \mathds{1}~.
\end{equation}
Hence, the state of the old black hole is maximally mixed. We stress that this is an approximate statement that does not hold in the exact theory. However, for the purpose of information recovery at late times this seems to be sufficient. 

We define $\mathcal{N}$ to be the algebra of operators acting on the black hole Hilbert space $\mathcal{H}_{M}$. Note that this convention is different from the one used in section \ref{sec:2} where the algebra $\mathcal{N}$ was defined on the Hilbert space $\mathcal{H}_{M-E}$. However, to avoid writing the subscript $M-E$ everywhere we have conveniently shifted the mass by $E$. The corresponding operators constitute a type I$_{d_{M}}$ von Neumann algebra, so we can identify $\mathcal{N}$ with the matrix algebra $M_{d_M}(\mathbb{C})$ of $d_{M}\times d_{M}$ matrices that act on the black hole microstates. The GNS Hilbert space associated to the algebra $\mathcal{N}$ is now simply given by a subspace 
\begin{equation} 
\mathcal{H}_{\mathcal{N}}\equiv \mathcal{H}_M \otimes \mathcal{H}_M' \subset \mathcal{H}_{M} \otimes \mathcal{H}_{R}~,
\end{equation}
where $\mathcal{H}_M'$ is a copy of the Hilbert space $\mathcal{H}_M$, and the embedding $\mathcal{H}_{M}'\subset \mathcal{H}_R$ describes that part of the old radiation that is entangled with the black hole. The orthogonality relation \eqref{eq:orthogonal} implies that $|\chi\rangle$ is to good approximation a maximally entangled state, and thus cyclic and separating for the algebra $\mathcal{N}$ in the Hilbert space $\mathcal{H}_{\mathcal{N}}$. 

\subsection{Introducing Alice's diary}

Let us now include a Hilbert space $\mathcal{H}_A$ that can be used to store the information of Alice's diary. Its dimension is denoted by $d_A\equiv \mathrm{dim}\, \mathcal{H}_A$, and the states in the Hilbert space are labeled by $|\alpha\rangle$. We think about the combined system $\mathcal{H}_{M}\otimes \mathcal{H}_{A}$ as a somewhat larger black hole, where we have identified part of the interior with the diary that has fallen in. Alice can act with arbitrary operators on this part of the Hilbert space, and we denote this algebra by $\mathcal{A}$. We also define the algebra 
\begin{equation} \label{eq:M=NxA}
\mathcal{M} \equiv \mathcal{N}\otimes \mathcal{A}~,
\end{equation}
that acts on the combined system of the original black hole and the diary Hilbert space. The corresponding GNS Hilbert space $\mathcal{H}_{\mathcal{M}}$ takes the form 
\begin{equation} \label{eq:GNSM}
\mathcal{H}_{\mathcal{M}} \equiv \mathcal{H}_{M}\otimes \mathcal{H}_A \otimes \mathcal{H}_M'\otimes  \mathcal{H}_A'~,
\end{equation}
where $\mathcal{H}_A'$ is an extra copy of the diary Hilbert space, that we can think of as a way to purify mixed states on $\mathcal{H}_A$. We distinguish between the states in $\mathcal{H}_A$ and $\mathcal{H}_A'$ by writing a primed label for the states in $\mathcal{H}_A'$, i.e., $|\alpha'\rangle \in \mathcal{H}_A'$. When we define the commutant of the subalgebra $\mathcal{N}$ with respect to the Hilbert space $\mathcal{H}_{\mathcal{M}}$, it is easy to see from the tensor product structure in \eqref{eq:M=NxA} that Alice's operators are indeed given by the relative commutant 
\begin{equation} 
	\mathcal{A} = \mathcal{N}'\cap \mathcal{M}~.
\end{equation}
Since we are dealing with finite-dimensional matrix algebras, we can write the index of the inclusion $\mathcal{N}\subset \mathcal{M}$ in terms of the dimension of the diary Hilbert space. It is simply given by the ratio of dimensions of the corresponding GNS Hilbert spaces
\begin{equation} 
	[\mathcal{M}:\mathcal{N}]=\frac{\dim \mathcal{H}_{\mathcal{M}}}{\dim \mathcal{H}_{\mathcal{N}}}=d_{A}^2~,
\end{equation} 
and therefore it provides a direct measure for the size of the diary Hilbert space.
\begin{figure}
	\centering
	\includegraphics[width=0.45\linewidth]{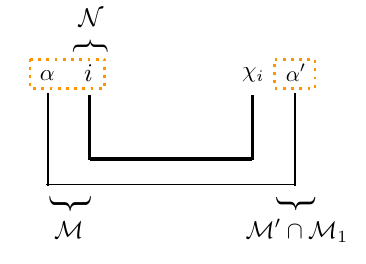}
	\caption{A schematic depiction of the state $|\Phi\rangle$ in the Hilbert space $\mathcal{H}_{\mathcal{M}}=\mathcal{H}_{M}\otimes \mathcal{H}_A \otimes \mathcal{H}_M'\otimes  \mathcal{H}_A'$. We have indicated on which part of the Hilbert space the algebras $\mathcal{N}, \mathcal{M}$ and $\mathcal{M}_1$ act. In particular, the indices corresponding to the basic extension algebra $\mathcal{M}_1=J_{\mathcal{M}}\mathcal{N}'J_{\mathcal{M}}$ are enclosed by a colored box. Note that $\mathcal{M}_1$ has non-trivial overlap with the commutant $\mathcal{M}'$.}
	\label{fig:algebrasm1}
\end{figure}

We denote the maximally entangled state on the combined system of the black hole and the Hilbert space of Alice's diary by 
\begin{equation} \label{eq:Phi}
|\Phi \rangle \equiv \frac{1}{\sqrt{d_Md_A} }\sum_{i,\alpha}|i,\alpha \rangle |\chi_i, \alpha '\rangle~.
\end{equation}
An operator $a\in \mathcal{M}$ acts on the indices $i$ and $\alpha$, so when represented on the GNS Hilbert space $\mathcal{H}_{\mathcal{M}}$ it should be identified with $a\otimes 1$. By the cyclic property of $|\Phi\rangle$ a general state in $\mathcal{H}_{\mathcal{M}}$ can now be expressed as
\begin{equation} \label{eq:generatorsHM}
|a\rangle \equiv a \otimes 1|\Phi\rangle = \frac{1}{\sqrt{d_Md_A}}\sum_{i,\alpha,j,\beta}a_{i\alpha,j\beta}|j,\beta\rangle|\chi_i,\alpha'\rangle~, \quad a\in \mathcal{M}~,
\end{equation}
in terms of the matrix elements $a_{i\alpha,j\beta}$ of the operator $a$. The smaller algebra $\mathcal{N}$ only acts on the index $i$, and does nothing to the index $\alpha$. For this reason, the subspace $\mathcal{H}_{\mathcal{N}}$ consists of states
\begin{equation} 
|b\rangle \equiv b \otimes 1|\Phi\rangle = \frac{1}{\sqrt{d_Md_A}}\sum_{i,\alpha,j}b_{i,j}|j,\alpha\rangle|\chi_i,\alpha'\rangle~, \quad b\in\mathcal{N}~.
\end{equation}
From the above expressions it is clear how to define a projector $e_{\mathcal{N}}$ that maps $\mathcal{H}_{\mathcal{M}}$ onto $\mathcal{H}_{\mathcal{N}}$. Given a general state \eqref{eq:generatorsHM}, one simply has to identify the labels $\beta$ and $\alpha$. This can be achieved by applying the projection
\begin{equation} \label{eq:projectorA}
e_{\mathcal{N}}=|\psi\rangle_A \langle \psi |_A~,
\end{equation}
onto the state $|\psi\rangle_A$ that represents the maximally entangled Bell pair shared between $\mathcal{H}_A$ and $\mathcal{H}_{A}'$,
\begin{equation} \label{eq:BellstateA}
|\psi\rangle_A \equiv \frac{1}{\sqrt{d_A}}\sum_{\alpha}|\alpha\rangle |\alpha'\rangle~. 
\end{equation}
In that sense, the projector $e_{\cal N}$ entangles the subalgebras corresponding to $\mathcal{H}_A$ and $\mathcal{H}_{A}'$. Using the expression \eqref{eq:generatorsHM} it is easy to verify that the above projection operator acts via
\begin{equation} 
e_{\mathcal{N}}|a\rangle = |\mathcal{E}(a)\rangle~,
\end{equation}
where the map $\mathcal{E}:\mathcal{M} \to \mathcal{N}$ is given by a partial trace over the diary subsystem
\begin{equation} \label{eq:condexptypei}
\mathcal{E}(a)=\frac{1}{d_A}\mathrm{tr}_{A}(a)=\frac{1}{d_A}\sum_{\alpha}(1\otimes \langle \alpha|) a (1\otimes |\alpha\rangle)~.
\end{equation}
Equation \eqref{eq:condexptypei} defines a conditional expectation that removes the algebra $\mathcal{A}$ from $\mathcal{M}$.

It is important to observe that the projector $e_{\mathcal{N}}$ is not part of the algebra $\mathcal{M}$, as it acts on the index $\alpha'$. Therefore, adding the  projector to the algebra leads to a larger von Neumann algebra that we denote by 
\begin{equation} 
\mathcal{M}_1\equiv \langle \mathcal{M}, e_{\mathcal{N}} \rangle~. 
\end{equation}
It is precisely the basic extension algebra that acts on the states in $\mathcal{H}_{M}\otimes \mathcal{H}_A$ as well as the states in $\mathcal{H}_A'$. It is useful to have a pictorial representation of the above definitions in mind, see Figure \ref{fig:algebrasm1}. 

As was explained in section \ref{sec:2}, there is an equivalent description of the algebra $\mathcal{M}_1$ in terms of the modular conjugation operator. The action of $J_{\mathcal{M}}$ depends on the specific entanglement structure that is present in the state $|\Phi\rangle$. It follows by a standard computation that for a maximally entangled state $J_{\mathcal{M}}$ acts as a ``swap operator'' on the basis elements, in the sense that it swaps the labels of the unprimed and corresponding primed indices,
\begin{equation} \label{eq:explicitJM}
J_{\mathcal{M}}: |i,\alpha\rangle|\chi_j,\beta'\rangle \mapsto |j,\beta\rangle |\chi_{i}, \alpha'\rangle~.
\end{equation}
Using the explicit form of the modular conjugation it one can easily check that 
\begin{equation} 
\mathcal{M}_1 =J_{\mathcal{M}}\mathcal{N}'J_{\mathcal{M}}~,
\end{equation}
where $\mathcal{N}'$ is the commutant of $\mathcal{N}$ within $\mathcal{H}_{\mathcal{M}}$. In Figure \ref{fig:algebrasm1} the modular conjugation $J_{\mathcal{M}}$ acts via a reflection in the vertical axis. 

The entanglement in the Bell state \eqref{eq:BellstateA} can be used to transfer information. To be precise, we have the well-known property 
\begin{equation} \label{eq:pushingproperty}
a\otimes 1 |\psi\rangle_A = 1\otimes a^{T} |\psi\rangle_A~.
\end{equation}  
This is exactly the type I version of the identity \eqref{eq:gammaprojector} that was crucial to achieve information transfer in the type II$_1$ setting. To see this, we define the map
\begin{equation} 
\gamma_{\mathcal{M}}(a)\equiv J_{\mathcal{M}}a^{\dagger}J_{\mathcal{M}}~,
\end{equation}
as before. Recall that it maps Alice's operator algebra to
\begin{equation} 
a\in \mathcal{A} \hspace{10pt} \longrightarrow \hspace{10pt}  \gamma_{\mathcal{M}}(a) \in \mathcal{M}'\cap \mathcal{M}_1~.
\end{equation}
Using the explicit form of the modular conjugation in \eqref{eq:explicitJM}, one can indeed show that 
\begin{align}
\gamma_{\mathcal{M}}(a\otimes 1)|\psi\rangle_A &=\frac{1}{\sqrt{d_A}}\sum_{\alpha} J_{\mathcal{M}}(a^{\dagger}\otimes 1)|\alpha\rangle|\alpha'\rangle =\frac{1}{\sqrt{d_A}}\sum_{\alpha,\beta} J_{\mathcal{M}}(a_{\beta \alpha}^*|\beta \rangle|\alpha'\rangle) \nonumber \\
&=\frac{1}{\sqrt{d_A}}\sum_{\alpha,\beta} a_{\beta \alpha}J_{\mathcal{M}}|\beta \rangle|\alpha'\rangle = \frac{1}{\sqrt{d_A}}\sum_{\alpha,\beta} a_{\beta \alpha}|\alpha \rangle|\beta'\rangle = 1\otimes a^{T} |\psi\rangle_A~.
\end{align}
This is precisely the right-hand side of \eqref{eq:pushingproperty}.  

\subsection{Bob and the entanglement swap}

Importantly, the recovery protocol consists of an additional step. To this end, we introduce the GNS Hilbert space $\mathcal{H}_{\mathcal{M}_1}$ associated to the algebra $\mathcal{M}_1$ in our microscopic model. This adds another copy of the Hilbert space $\mathcal{H}_A$ that we will denote by $\mathcal{H}_B$, as it will be used by Bob to extract the information. It has the same dimension $d_B \equiv \dim \mathcal{H}_B$ as the Hilbert space $\mathcal{H}_A$, and with slight abuse of notation we will use the same set of basis states $|\alpha\rangle$ to describe it. The GNS Hilbert space can now be represented as
\begin{equation} \label{eq:tensorhilbert}
\mathcal{H}_{\mathcal{M}_1} \equiv \mathcal{H}_{M}\otimes \mathcal{H}_A \otimes  \mathcal{H}_A' \otimes \mathcal{H}_M' \otimes \mathcal{H}_B'\otimes \mathcal{H}_B~.
\end{equation}
Comparing this expression with \eqref{eq:GNSM}, one has to keep in mind that within the Hilbert space $\mathcal{H}_{\mathcal{M}}$ the algebra $\mathcal{M}_1$ is part of the commutant $\mathcal{M}'$, as it acts on the Hilbert space $\mathcal{H}_{A'}$. However, when representing the action of $\mathcal{M}_1$ on the GNS representation of $\mathcal{M}_1$ we group the Hilbert space $\mathcal{H}_{A}\otimes \mathcal{H}_{A}'$ to the left and the newly introduced Hilbert space $\mathcal{H}_B'\otimes \mathcal{H}_B$, that can be used to purify states in $\mathcal{H}_{A}\otimes \mathcal{H}_{A}'$, to the right.

\begin{figure}
	\centering
	\includegraphics[width=0.6\linewidth]{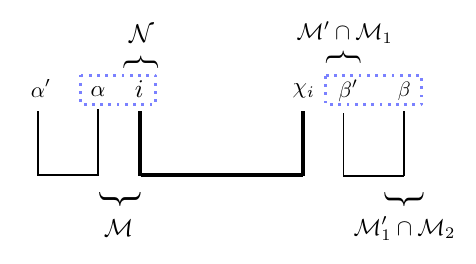}
	\caption{A schematic depiction of the state $|\Psi\rangle$ in the Hilbert space $\mathcal{H}_{\mathcal{M}_1}=\mathcal{H}_{M}\otimes \mathcal{H}_A \otimes  \mathcal{H}_A' \otimes \mathcal{H}_M' \otimes \mathcal{H}_B'\otimes \mathcal{H}_B$. The modular conjugation map $J_{\mathcal{M}_1}$ acts by swapping the indices that are connected by lines. The indices corresponding to the algebra $\mathcal{M}_2=J_{\mathcal{M}_1}\mathcal{M}'J_{\mathcal{M}_1}$ are enclosed by a colored box. We have also indicated where the relative commutants $\widetilde{\mathcal{B}}=\mathcal{M}'\cap \mathcal{M}_1$ and $\mathcal{B}=\mathcal{M}_1'\cap \mathcal{M}_2$ act.}
	\label{fig:gnsm1}
\end{figure}

Let us denote by $|\Psi\rangle$ the maximally entangled state on $\mathcal{H}_{\mathcal{M}_1}$. In terms of basis states, it can be written out explicitly as
\begin{equation} \label{eq:Psi}
|\Psi\rangle = \frac{1}{\sqrt{d_Md_Ad_B} }\sum_{i,\alpha,\beta}|i,\alpha, \beta' \rangle |\chi_i, \alpha' , \beta\rangle~,
\end{equation}
where the decomposition into left and right states is according to \eqref{eq:tensorhilbert},
\begin{equation} \label{eq:tensorfactors}
|i,\alpha, \beta' \rangle \in \mathcal{H}_{M}\otimes \mathcal{H}_{A}\otimes \mathcal{H}_{A}'~, \quad |\chi_i,\alpha', \beta \rangle \in \mathcal{H}_{M}'\otimes \mathcal{H}_{B}'\otimes \mathcal{H}_{B}~,
\end{equation}
We will now introduce the modular conjugation operator $J_{\mathcal{M}_1}$, and implement the basic construction a second time. The map $J_{\mathcal{M}_1}$ is defined with respect to the state $|\Psi\rangle$ on $\mathcal{H}_{\mathcal{M}_1}$. Again, it swaps the labels of the basis states, but now with respect to the entanglement structure of \eqref{eq:Psi}. To be precise, we have 
\begin{equation} 
J_{\mathcal{M}_1}:|i,\alpha, \beta' \rangle |\chi_j, \gamma' , \delta \rangle \mapsto |j,\gamma, \delta' \rangle |\chi_i, \alpha' , \beta\rangle~.
\end{equation}

As in section \ref{sec:2}, the algebra $\mathcal{M}_2$ can now be obtained from the commutant of $\mathcal{M}$ within the GNS Hilbert space of $\mathcal{M}_1$ by 
\begin{equation} 
\mathcal{M}_2 =J_{\mathcal{M}_1}\mathcal{M}'J_{\mathcal{M}_1}~.
\end{equation} 
Similarly, we define Bob's radiation algebra as the relative commutant
\begin{equation} 
\mathcal{B}\equiv \mathcal{M}_1'\cap \mathcal{M}_2~,
\end{equation}
which corresponds to the operators that act on the Hilbert space $\mathcal{H}_B$. The state $|\Psi\rangle$ and the relevant algebras are depicted in Figure \ref{fig:gnsm1}. As before, defining the map 
\begin{equation} 
\gamma_{\mathcal{M}_1}(a)\equiv J_{\mathcal{M}_1}a^{\dagger}J_{\mathcal{M}_1}~,
\end{equation} 
the canonical shift $\Gamma:\mathcal{A}\to \mathcal{B}$ can be expressed in terms of the composition
\begin{equation} 
\Gamma = \gamma_{\mathcal{M}_1}\circ \gamma_{\mathcal{M}}: a \mapsto J_{\mathcal{M}_1}J_{\mathcal{M}}aJ_{\mathcal{M}}J_{\mathcal{M}_1}~.
\end{equation}
Looking at the tensor product structure of the Hilbert space \eqref{eq:tensorhilbert} we see that the canonical shift acts in two steps,
\begin{equation} \label{eq:entanglementswap}
\mathcal{A} \otimes 1_{A'} \otimes 1_{B} \otimes 1_{B'} \to  1_{A} \otimes \mathcal{A} \otimes 1_{B} \otimes 1_{B'} \to 1_{A} \otimes 1_{A'} \otimes  \mathcal{A} \otimes 1_{B'}~,
\end{equation}
making use of the intermediate channel $\mathcal{H}_A'$.

The type I setting makes manifest that the entanglement structure of the global state plays an important role in transferring an operator from $\mathcal{A}$ to $\mathcal{B}$. The canonical shift amounts to two consecutive swaps: In the first step the entanglement between the system $A$ and $A'$ in the state $|\Phi\rangle$ is used to transfer the operator acting on the system $A$ to $A'$, while in the second step, we use the entanglement between $A'$ and $B$ in the state $|\Psi\rangle$ to transfer the operator into the algebra $\mathcal{B}$. In particular, we see that an \emph{entanglement swap} needs to take place. In the first step, we are in a situation where $A$ and $A'$ are entangled with each other, while in the second step the entanglement with $A$ has been transferred to $B$. This entanglement swap is at the heart of the recovery protocol, and provides some explanation for why the smooth horizon and unitarity can coexist, see also \cite{Horowitz:2003he, Lloyd:2013bza, Yoshida:2019qqw}.

To see that the information got transferred correctly, one needs to do a similar computation as in section \ref{sec:3}. As before, the state of the system after Alice has thrown her diary into the black hole is given by 
\begin{equation} 
|\vxi\rangle \equiv \vxi|\Psi\rangle = \frac{1}{\sqrt{d_Md_Ad_B}} \sum_{i,\alpha,\beta, \gamma}\vxi_{\alpha\gamma}|i,\gamma,\beta'\rangle|\chi_i,\alpha',\beta\rangle~,
\end{equation}
where $\vxi\in \mathcal{A}$. Recall that Bob has recovered the information of the diary once he can reproduce from his algebra $\mathcal{B}$ correlation functions of the form
\begin{equation} \label{eq:explicitcorrfunction} 
\langle \vxi|a| \vxi \rangle =\mathrm{tr}_A(a\vxi \vxi^{\dagger})~,  \qquad \mathrm{for} \qquad a\in \mathcal{A}~.
\end{equation}  
\begin{figure}
	\centering
	\includegraphics[width=0.6\linewidth]{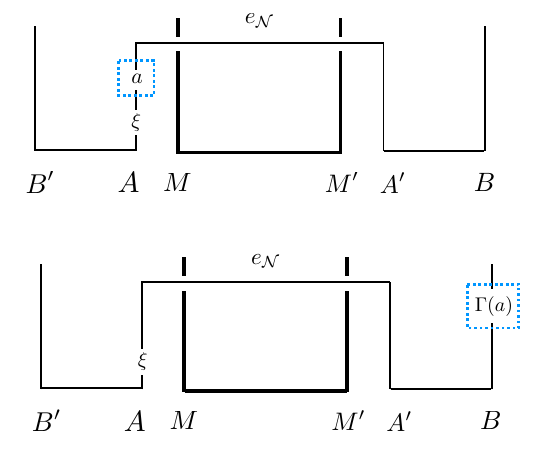}
	\caption{A teleportation step. After applying the projector $e_{\mathcal{N}}$ to the state $|\vxi\rangle$ Bob can teleport an operator in Alice's algebra to his own algebra using the canonical shift $a \in \mathcal{A} \mapsto \Gamma(a)\in \mathcal{B}$.}
	\label{fig:teleportationstep}
\end{figure}
Associated to each $a\in \mathcal{A}$ there is a canonical operator $\Gamma(a)\in \mathcal{B}$ that Bob can use, and the information of the operator is indeed transferred once he has applied $e_{\mathcal{N}}$ to the state $|\vxi\rangle$. The projector $e_{\mathcal{N}}$ in \eqref{eq:projectorA} identifies the states in $\mathcal{H}_A$ and $\mathcal{H}_{A'}$ by projecting onto a maximally entangled Bell pair, so the projected state $e_{\mathcal{N}}|\vxi\rangle$ can be represented as
\begin{equation} 
e_{\mathcal{N}}|\vxi\rangle =  \frac{1}{d_A\sqrt{d_Md_Ad_B}} \sum_{i,\alpha,\beta, \gamma}\vxi_{\alpha\beta}|i,\gamma,\gamma'\rangle|\chi_i,\alpha',\beta\rangle~.
\end{equation}
Let us now act with the operator $\Gamma(a)$ on the above state. It is clear that 
\begin{equation} 
J_{\mathcal{M}}J_{\mathcal{M}_1}|i,\gamma,\gamma'\rangle|\chi_i,\alpha',\beta\rangle = J_{\mathcal{M}}|i,\alpha,\beta'\rangle|\chi_i,\gamma',\gamma\rangle = |i,\beta,\alpha'\rangle|\chi_i,\gamma',\gamma\rangle~.
\end{equation}
Acting with $a$ on the above state, and then applying $J_{\mathcal{M}_1}J_{\mathcal{M}}$ again we find that 
\begin{equation} 
\Gamma(a)e_{\mathcal{N}}|\vxi\rangle = \frac{1}{d_A\sqrt{d_Md_Ad_B}} \sum_{i,\alpha,\beta, \gamma,\delta}\vxi_{\alpha\beta}a_{\beta\delta}|i,\gamma,\gamma'\rangle|\chi_i,\alpha',\delta\rangle~.
\end{equation}
This makes explicit that the canonically shifted operator $\Gamma(a)$ acts on the last tensor factor in \eqref{eq:tensorfactors}, which is precisely the Hilbert space $\mathcal{H}_B$. Using the normalized state 
\begin{equation} 
|\vxi\rangle_{B} \equiv d_A e_{\mathcal{N}} |\vxi\rangle~,
\end{equation}
Bob therefore computes the following correlation function
\begin{equation} 
\langle \vxi|_B\Gamma(a)|\vxi\rangle_B =\frac{1}{d_A} \sum_{\alpha,\beta,\gamma}\vxi^*_{\alpha\gamma}\vxi_{\alpha\beta}a_{\beta\gamma}=\mathrm{tr}_A(a\vxi\vxi^{\dagger})~,
\end{equation}
which agrees with the expression in \eqref{eq:explicitcorrfunction}. Pictorially, the information transfer is given in Figure \ref{fig:teleportationstep}.

\subsection{The dynamics of the black hole: a scrambling unitary}

We have seen that Bob can reconstruct the information of Alice's diary when he is in possession of a suitable conditional expectation. This might seem like a very strong requirement: To be able to project out the information of the diary he essentially needs to know in what part of the black hole Hilbert space the information is contained.  We will now argue that the situation is actually much better and Bob can use the newly emitted Hawking radiation to define such a conditional expectation provided that the black hole dynamics is sufficiently scrambling. 

Up to this point we have somewhat oversimplified the set-up, and completely neglected the dynamics of the black hole in our type I description. Of course, this is an essential feature of the problem and should be included in a more complete model of the evaporation process. We assume that the dynamics of the black hole in the time step between $t_A$ and $t_B$ is given by some unitary \begin{equation} 
U: \mathcal{H}_M\otimes \mathcal{H}_A\to \mathcal{H}_{\widetilde{M}}\otimes \mathcal{H}_{\rm rad} ~.
\end{equation}
This operation describes both the scrambling of the diary with the black hole degrees of freedom and the subsequent emission of additional Hawking particles. We have introduced the Hilbert space  $\mathcal{H}_{\rm rad}$ and $\mathcal{H}_{\widetilde{M}}$ to describe the new Hawking quanta that have come out and the remaining black hole degrees of freedom. At time $t=t_B$ the state of the system is given by
 \begin{equation}
 U\otimes 1 | \vxi\rangle  ~.
 \end{equation}
We assume that Bob has access to the algebra $\mathcal{M}'$ that acts on the old radiation entangled with the original black hole and the auxiliary system $A'$. As he is familiar with the precise form of the black hole dynamics he can construct a unitary $U^{\dagger}$ that undoes the action of $U$. Moreover, having obtained the additional Hawking quanta he can construct the following projector,  
\begin{equation}
e_{\rm rad} = |\psi\rangle_{\rm rad} \langle \psi|_{\rm rad}~, \hspace{20pt}  |\psi\rangle_{\rm rad} \equiv \frac{1}{d_{\rm rad}}\sum_{r}|r\rangle|r\rangle~,
\end{equation}
that corresponds to the purification of the state of the new Hawking particles. To retrieve the information he needs to apply the following operator, 
\begin{equation}
V =e_{\rm rad} J_{\mathcal{M}}(U^{\dagger}\otimes 1) J_{\mathcal{M}}
\end{equation}
to the state $ U\otimes 1 | \vxi\rangle$. Bob's state  after applying $V$ is given by 
\begin{equation} 
	|\vxi\rangle_{B} \equiv d_{\rm rad} V (U\otimes 1) |\vxi\rangle = d_{\rm rad} e_{\rm rad}(U\otimes U^{*})|\vxi\rangle~.
\end{equation}

\begin{figure}
	\centering
	\includegraphics[width=\linewidth]{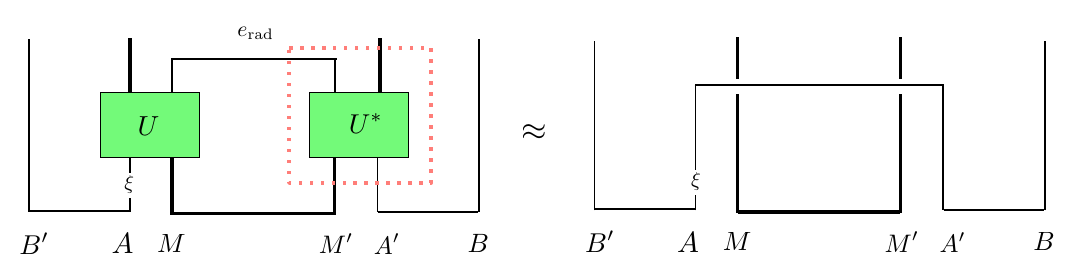}
	\caption{The role of scrambling in the teleportation protocol. If the unitary $U$ is sufficiently scrambling Bob can apply the operator $V=e_{\rm rad}(1\otimes U^*)$ (enclosed by the red box) to project out the information in Alice's Hilbert space. He can therefore use a projector $e_{\rm rad}$ constructed from the outgoing Hawking quanta to obtain, to good approximation, the state on the right-hand side. This retrieves the information.}
	\label{fig:scramblingu}
\end{figure}

The canonical shift $a\mapsto \Gamma(a)$ can now be used to retrieve the information provided that the unitary $U$ is sufficiently scrambling. This is exactly the  reasoning that goes into the Yoshida-Kitaev protocol \cite{Yoshida:2017non}. In fact, they give necessary and sufficient conditions for the information to be retrievable: It is shown there that $(U\otimes U^*)^{\dagger}e_{\rm rad}(U\otimes U^*)$ is a good approximation to $e_{\mathcal{N}}$, when the size of the radiation subspace $d_{\rm rad}$ is somewhat larger than the dimension $d_{A}$ of the subspace associated to the diary. The error can be expressed in terms of the ratio $d_A/d_{\rm rad}$ \cite{Yoshida:2017non}. To summarize, Bob has to wait long enough so that enough Hawking pairs have come out, but provided that the dynamics of the black hole is sufficiently scrambling, he can to good approximation apply the projector $e_{\mathcal{N}}$. What information Bob can retrieve is precisely determined by what part of the Hilbert space is scrambled with respect to the Hawking particles that Bob uses to construct the projector $e_{\rm rad}$. 

\section{Towards an Emergent Spacetime Description}

\label{sec:5}

The ultimate goal of our work is to have a spacetime interpretation of the information recovery protocol, and make a connection with the experience of the in-falling observer. Understanding this would be an important step towards solving the black hole information problem as it amounts to reconciling information recovery with the existence of a smooth horizon. In the present paper, we will not solve this problem. We do believe, however, that the algebraic approach makes manifest certain features of the recovery protocol that are amenable to an emergent spacetime interpretation. We will now provide some evidence for this statement by observing that the canonical shift is still well-defined in the context of type III$_1$ algebras. Furthermore, in the case of a half-sided modular inclusion, which is the relevant algebraic set-up for describing the experience of the in-falling observer, the canonical shift has a clear and rather interesting physical interpretation: It corresponds to a spacetime translation.  We plan to give a more detailed account of these ideas in upcoming work \cite{workinprogress1}. 

\subsection{The canonical shift as a modular translation}

Let us switch gears and consider an inclusion $\mathcal{N}\subset \mathcal{M}$ of type III$_1$ von Neumann algebras with common cyclic and separating state $|\Omega\rangle$. Since we are considering type III von Neumann algebras, there is no preferred state in the form of the tracial state that exists for type II$_1$ von Neumann algebras. For this reason, the modular operator $\Delta_{\cm}$ associated to the state $|\Omega\rangle$ is generically non-trivial, and defines a modular time flow on the algebra $\cm$ via the assignment
\beq \label{eq:modularflow1}
\sigma_s(a)\equiv \Delta_{\mathcal{M}}^{-is}a\Delta_{\mathcal{M}}^{is}~.
\eeq
We can now consider a subalgebra $\cn \subset \cm$ such that the modular flow of the algebra $\cm$ maps $\cn$ into itself for positive modular times, 
\beq \label{eq:HSM}
\sigma_s(\mathcal{N})\subset \mathcal{N}~, \hspace{20pt} \mathrm{for} \hspace{20pt} s\geq 0~.
\eeq 
This situation is called a half-sided modular inclusion \cite{Wiesbrock:1992mg}. An important property of a half-sided modular inclusion, which makes it a rather powerful, is the existence of a one-parameter group of unitaries that can be used to translate the algebra $\cm$ into $\cn$ \cite{Wiesbrock:1992mg, Araki:2005we}. To define such a translation operator, we first need to introduce the modular Hamiltonians $K_{\cm}, K_{\cn}$ by the relation
\begin{equation} 
\Delta_{\cm}=e^{-2\pi K_{\cm}}~, \qquad \Delta_{\cn}=e^{-2\pi K_{\cn}}~,
\end{equation} 
where $\Delta_{\cm}, \Delta_{\cn}$ are the modular operators associated to $\cm$ and $\cn$ respectively. Using Borchers's theorem \cite{Borchers:1991xk}, the half-sided modular inclusion property implies that there exists a positive operator $P$ with the following properties
\begin{equation} \label{eq:defP}
P = K_{\cm}-K_{\cn}~, \qquad [K_{\cm},K_{\cn}]=i P~.
\end{equation}
We now have a one-parameter group of unitaries 
\begin{equation} 
U(s)=e^{itP}~, \hspace{10pt} t\in \mathbb{R}~,
\end{equation}
that can be used to translate the algebra. The existence of a generator $P$ with the above properties in the case of a half-sided modular inclusion is a very powerful result. For example, the operators in translated algebra do not necessarily commute with the commutant of the original algebra. In particular, the subalgebra $\mathcal{N}$ can be mapped onto $\mathcal{M}$ by conjugating with the unitary $U(1)$. We have 
\beq 
\mathcal{M} = U(1)\mathcal{N}U(-1)~.
\eeq
This was used in \cite{Leutheusser:2021qhd, Leutheusser:2021frk} to argue that the translation operator provides a natural time coordinate $t$ for the in-falling observer in the context of an eternal AdS black hole.

The present goal is to find an expression for the canonical shift in the case of a half-sided modular inclusion. To this end, we first derive the following useful relation for the translation operator
\begin{equation} \label{eq:identityP}
\Delta_{\mathcal{M}}^{1/2} \Delta_{\mathcal{N}}^{-1/2}=e^{-2 i P}~,
\end{equation}
which is a special case of the more general identity
\begin{equation} \label{eq:identityP2}
e^{2 \pi i t K_{\mathcal{M}}} e^{-2 \pi i t K_{\mathcal{N}}}=e^{i\left(e^{2 \pi t}-1\right) P}~.
\end{equation}
Although the statement in \eqref{eq:identityP2} is somewhat standard in the theory of half-sided modular inclusions, we have included a short derivation for completeness. It relies on the Baker-Campbell-Hausdorff formula. By differentiating the left-hand side of \eqref{eq:identityP2} we obtain
\begin{equation} \label{eq:d/dt}
\frac{d}{d t}\left(e^{2 \pi i t K_{\mathcal{M}}} e^{-2 \pi i t K_{\mathcal{N}}}\right)=2 \pi i\left(e^{2 \pi i t K_{\mathcal{M}}} P e^{-2 \pi i t K_{\mathcal{N}}}\right)~.
\end{equation}
Here, we have used the product rule combined with the first equation in \eqref{eq:defP}. The second equation in \eqref{eq:defP} can now be used to derive the commutation relation of $P$ with $K_{\mathcal{M}}$, which is given by 
\begin{equation}
\left[K_{\mathcal{M}}, P\right]=-i P~.
\end{equation}
With the help of this commutator we can rewrite the right-hand side of \eqref{eq:d/dt} by moving the operator $P$ to the left of the exponential. It follows that
\begin{equation} \label{eq:differentialeq}
\frac{d}{d t}\left(e^{2 \pi i t K_{\mathcal{M}}} e^{-2 \pi i t K_{\mathcal{N}}}\right)=2 \pi i e^{2 \pi t} P\left(e^{2 \pi i t K_{\mathcal{M}}} e^{-2 \pi i t K_{\mathcal{N}}}\right)~.
\end{equation}
It is convenient to introduce the function
\begin{equation}
F(P, t)\equiv e^{2 \pi i t K_{\mathcal{M}}} e^{-2 \pi i t K_{\mathcal{N}}}~,
\end{equation}
so that \eqref{eq:differentialeq} is given by the following differential equation:
\begin{equation}
\frac{d}{d t} F(P, t)=2 \pi i e^{2\pi t} P F(P, t) \quad \text { with } \quad F(P, 0)=1~.
\end{equation}
It is straightforward to check that the solution is given by
\begin{equation}
F(P, t)=e^{i\left(e^{2 \pi t}-1\right) P}~.
\end{equation}
This proves equation \eqref{eq:identityP2}, and by analytic continuation to $t=i / 2$ we also obtain equation \eqref{eq:identityP}. 

We now use this result to write the operator $P$ in terms of the modular conjugation operators. To this end, we decompose the Tomita operators associated to $\mathcal{M}$ and $\mathcal{N}$ in the following way
\beq 
S_{\mathcal{M}} = J_{\mathcal{M}}\Delta_{\mathcal{M}}^{1/2}~, \hspace{20pt} S_{\mathcal{N}} = \Delta_{\mathcal{N}}^{-1/2}J_{\mathcal{N}}~.
\eeq
Using these decompositions and the fact that the modular conjugation squares to one, see \eqref{eq:propJM}, we can rewrite the left-hand side of \eqref{eq:identityP} as
\beq 
\Delta_{\mathcal{M}}^{1/2}\Delta_{\mathcal{N}}^{-1/2}=J_{\mathcal{M}}S_{\mathcal{M}}S_{\mathcal{N}}J_{\mathcal{N}}~.
\eeq
We want to evaluate this operator on states of the form $|a\rangle =  a|\Omega\rangle$ with $a\in \mathcal{N}'$. We can use the previous equation to derive that 
\begin{align} \label{eq:DeltaJ}
\Delta_{\mathcal{M}}^{1/2}\Delta_{\mathcal{N}}^{-1/2}|a\rangle &= J_{\mathcal{M}}S_{\mathcal{M}}S_{\mathcal{N}}J_{\mathcal{N}}a|\Omega\rangle = J_{\mathcal{M}}S_{\mathcal{M}}(J_{\mathcal{N}}aJ_{\mathcal{N}})^{\dagger}|\Omega\rangle =J_{\mathcal{M}}J_{\mathcal{N}}|a\rangle~.
\end{align}
Combining \eqref{eq:identityP} and  \eqref{eq:DeltaJ}, we conclude that the translation over twice the distance between $\mathcal{N}$ and $\mathcal{M}$ can be written as the composition of two modular conjugations. In formulas, we have 
\beq \label{eq:JJP}
J_{\mathcal{M}}J_{\mathcal{N}}=e^{-2iP}~.
\eeq
We note that this identity is well-known in the literature, see for example the original paper by Wiesbrock \cite[Corollary 4]{Wiesbrock:1992mg}, and we have simply provided a derivation to be somewhat self-contained.

\begin{figure}
    \centering
    \includegraphics[width=0.40\linewidth]{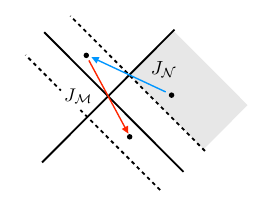}
    \caption{A geometric identity. We consider the algebras associated to two nested Rindler wedges $\mathcal{N}\subset \mathcal{M}$. The subregion associated to the subalgebra $\mathcal{N}$ is depicted in gray. It is a simple geometric argument to see that the composition of two modular conjugations $J_{\mathcal{M}}J_{\mathcal{N}}$ leads to a translation over twice the distance between the algebras $\mathcal{N}$ and $\mathcal{M}$.}
    \label{fig:JJPP}
\end{figure}

In the setting where $\mathcal{N}$ and $\mathcal{M}$ describe type III$_1$ von Neumann algebras associated to nested Rindler wedges, the identity \eqref{eq:JJP} takes a particularly intuitive form. The half-sided modular inclusion property \eqref{eq:HSM} is satisfied if the two wedges share a future null boundary, as depicted in Figure \ref{fig:JJPP}. It can be shown that in this case the modular conjugation acts geometrically via a point reflection in the bifurcate horizon, so that \eqref{eq:JJP} is easy to understand: It is simply the statement that applying two consecutive reflections over different bifurcation points leads to a translation over twice the distance between both bifurcation points.  

Let us now use the previous result to interpret the canonical shift in a spacetime setting. If we assume that the inclusion of algebras $\mathcal{N}\subset \mathcal{M}$ is a half-sided modular inclusion it follows that the algebra $\mathcal{M}_1$ is obtained from $\mathcal{N}$ by conjugating with the operator $U(-2)$. Using that $J_{\mathcal{M}_1}=J_{\mathcal{M}}J_{\mathcal{N}}J_{\mathcal{M}}$ one can show that the canonical shift can be expressed as
\beq 
\Gamma = J_{\mathcal{M}_1}J_{\mathcal{M}} = J_{\mathcal{M}}J_{\mathcal{N}}J_{\mathcal{M}}^2 =J_{\mathcal{M}}J_{\mathcal{N}}~,
\eeq
so we see that the canonical shift can also be represented via the right-hand side of \eqref{eq:JJP} in terms of the generator $P$. We conclude that the canonical shift in this case takes the form of a modular translation operator 
\beq 
\Gamma = \mathrm{Ad}\left(e^{-2iP}\right)~,
\eeq
over twice the distance between $\mathcal{N}$ and $\mathcal{M}$. We would therefore like to think about the canonical shift as a discrete version of the translation operator. 

Let us end this section with a two remarks. First, it is interesting to point out that one can naturally iterate the above construction to obtain an infinite sequence of inclusions
\beq \label{eq:Jonestower}
\mathcal{N} \subset \mathcal{M} \subset \Gamma(\mathcal{N})\subset \Gamma(\mathcal{M}) \subset \Gamma^2(\mathcal{N}) \subset \cdots,
\eeq
where at each step the algebra is shifted over a finite distance with the generator $P$ according to \eqref{eq:JJP}. For an inclusion of type II$_1$ factors the above sequence, that is obtained by either adding projectors or using the consecutive modular conjugations, is known as the Jones tower. It is suggestive to think about the above sequence of inclusions \eqref{eq:Jonestower} as a discrete model for spacetime, in the sense that the translation operator provides a continuous interpolation of the Jones tower. There is, however, an important difference with the type II$_1$ setting, namely that the type III algebras are represented on the same Hilbert space, and that the state $|\Omega\rangle$ is assumed to be cyclic and separating for both. 

Our second remark is about the relation between the emergent translation operator and certain ergodic properties of the underlying operator algebra. We would like to highlight some interesting work in this direction \cite{DeBoer:2019kdj, Gesteau:2023rrx, Ouseph:2023juq}, where the existence of spacetime symmetries close to a bifurcate horizon, specifically the translation operators, are understood from the perspective of an ergodic hierarchy for the von Neumann algebra. The reason that the modular translation operators appear in this context is that they generate a maximal Lyapunov coefficient with the modular Hamiltonian, in the sense of \cite{DeBoer:2019kdj}. It is known that a useful tool for studying these modular scrambling modes is provided by the modular Berry phase \cite{Czech:2017zfq, Czech:2019vih} that one can associate to a one-parameter family of modular Hamiltonians (see also \cite{deBoer:2021zlm, Czech:2023zmq}). Since this geometric phase can also be abstractly defined in terms of the underlying von Neumann algebras \cite{workinprogress2}, we expect that it will be useful in understanding the emergent properties of spacetime, see also \cite{Chen:2022nwf,Banerjee:2023eew,Aalsma:2024qnf}.

\section{Discussion}
\label{sec:6}

In the present paper, we have presented a protocol for recovering black hole information in terms of the underlying operator algebras. This allowed us to understand information recovery, not only in a simplified finite-dimensional model of the evaporating black hole, but also in a setting where the observables constitute an infinite-dimensional type II$_1$ von Neumann algebra. In the latter case, there is a notion of relative size between infinite-dimensional subalgebras that stays finite as encoded by the index, and a corresponding version of the information paradox that is solved by our protocol. 

On the level of the algebras the information transfer is implemented by the canonical shift. Microscopically, the canonical shift can be interpreted as a quantum teleportation step. By performing an entanglement swap the information in the interior of the black hole is transferred to the outgoing Hawking particles. In the type II$_1$ setting the canonical shift also appears naturally in the Jones theory of subfactor inclusions where it acts as a shift on the level of relative commutants. Importantly, we have shown that the algebraic description allows us to make contact with a spacetime picture as well, since the composition of modular conjugations that appears in the definition of the canonical shift still makes sense for an inclusion of type III$_1$ algebras. In the specific case of a half-sided modular inclusion, it exactly corresponds to a translation operator that shifts the horizon inwards, allowing the information to escape from the black hole. 

Let us now discuss some subtleties and directions for future research. 

\subsection{Translation = teleportation}

It is tempting to speculate that the above ideas provide a microscopic explanation for the ``island region'' that covers part of the interior of the black hole and is used to purify the outside radiation. From our perspective, information that enters the island is transferred to the radiation system via a quantum teleportation step, that is algebraically implemented by the canonical shift. In the spacetime description, we expect that the information transfer leads to an actual shift of the QES in the sense that quantum extremal surfaces at different times are related by a translation along the horizon, in the $G\to 0$ limit. To motivate this, we can take some inspiration from the proposed relation between traversable wormholes and quantum teleportation \cite{Gao:2016bin, Susskind:2017nto} (see also  \cite{deBoer:2018ibj, DeBoer:2019yoe}). We expect similar reasoning to hold in our case, in the sense that Bob's application of the conditional expectation, which should play the role of the double trace deformation of Gao-Jafferis-Wall \cite{Gao:2016bin}, gives rise to a negative energy shockwave in spacetime that opens up a wormhole allowing Alice's diary to enter into the radiation system. We would also like to mention \cite{Jefferson:2018ksk} where some related ideas on interior reconstruction in the context of half-sided modular inclusions is discussed.

On a more fundamental level, the statement that the composition of consecutive modular conjugations can be related to a translation operator in the case of a half-sided modular inclusion, while it corresponds to a teleportation step in the microscopic model seems to be quite deep. We take this as a hint that the microscopic origin of the translation operator, that shifts the information into the radiation system, is intimately related to the teleportation protocol. It is tempting to speculate that this is more generally true, and that the act of movement within an emergent spacetime relies on an underlying mechanism of quantum teleportation. We will refer to this proposal as ``translation = teleportation''. This is very much in line with the expectation of ER = EPR \cite{VanRaamsdonk:2010pw, Maldacena:2013xja} which says that the connectedness of spacetime relies on quantum entanglement in the underlying theory. It is precisely this entanglement that is used in the teleportation step. 

To make these statements more precise requires a better understanding of how our information recovery protocol is to be interpreted in an emergent spacetime picture. We would like to stress that this is by no means straightforward. In making contact with the spacetime description it would be useful to have a more general version of the index that can also be defined in the context of type III von Neumann algebras. There exists a generalization of the index to arbitrary factors due to Kosaki \cite{Kosaki:1992zz, kosaki1998type} that relies on Haagerup's theory of operator valued weights \cite{haagerup1979operatorI,haagerup1979operatorII} and the spatial theory of Connes \cite{connes1980spatial}. However, even using this more general definition it turns out that a half-sided modular inclusion does not allow for conditional expectations with finite index\footnote{We would like to thank Antony Speranza for illuminating discussions on this point.}. This mathematical obstruction therefore seems to prevent a straightforward implementation of the recovery protocol, as presented in this work,  to the emergent spacetime description.  We postpone further discussion of this issue, and its possible resolution, to future work. 

\subsection{Double-scaled SYK and quantum groups}

Arguably, the ideas that are presented in the paper are phrased rather abstractly in terms of properties of the operator algebra in question. This is done on purpose to make certain general features of the approach manifest. It would, however, be illuminating to apply our formalism in an explicit physical model that exhibits a type II$_1$ von Neumann algebra. One such model that seems particularly promising is the double-scaled SYK (DSSYK) model. Notably, it has been solved exactly in terms of a diagrammatic procedure involving so-called chord diagrams \cite{Berkooz:2018jqr}. The way the chords are contracted follows from the Wick contraction rules that arise in the averaging over the random couplings. One can show that each intersection in the diagram contributes a factor $q=e^{-\lambda}$, where $\lambda$ is a parameter that defines the double-scaling limit $N,p \to \infty$ in the sense that the ratio $\lambda \sim p^2/N$ is kept fixed. Importantly, the correlation functions that one obtains can be reorganized into representations of a certain quantum group with deformation parameter $q$.

It is interesting to point out that a similar structure also arises in the abstract theory of type II$_1$ subfactors. Given an inclusion $\mathcal{N}\subset \mathcal{M}$ of type II$_1$ factors one can repeat the basic construction indefinitely to obtain an infinite sequence of algebras that is called the Jones tower
\begin{equation} \label{eq:Jonestower2}
\mathcal{N}\subset \mathcal{M}\subset \mathcal{M}_1 \subset \mathcal{M}_2 \subset \cdots ~,
\end{equation}
and a corresponding set of Jones projectors $e_{0}=e_{\mathcal{N}}, e_{1}=e_{\mathcal{M}}, e_2=e_{\mathcal{M}_1}, \ldots$ Crucially, it was shown that the projectors satisfy the following commutation relations 
\begin{equation} \label{eq:TL}
e_{i}e_{i\pm 1}e_{i} = \alpha e_{i}~, \hspace{20pt} e_i e_j = e_j e_i~, \hspace{10pt} |i-j|>1~.
\end{equation}
where the parameter $\alpha$ is related to the inverse of the index of the inclusion. The relations \eqref{eq:TL} define the so-called Temperley-Lieb algebra, which have a diagrammatic representation that is quite similar to the chord diagrams. In particular, one can simplify a complicated Temperley-Lieb diagram using \eqref{eq:TL}. The representations of the Temperley-Lieb algebra can be related to representations of the braid group. As a consequence, there is also a natural quantum group structure associated to the construction, where the deformation parameter $q$ is related to the index.

It would be interesting to see if the chord diagrams that appear in the solution of the DSSYK model have a more intrinsic meaning in terms of the underlying type II$_1$ algebra. Is there an interesting inclusion of subfactors in the operator algebra of DSSYK that one can study? And how should we interpret the canonical shift in this case? Moreover, we would like to investigate if there is a more direct relation between the quantum group structure that arises in the DSSYK model, the corresponding chord diagrammatics and the Temperley-Lieb algebra that arises in the Jones subfactor theory applied to the underlying type II$_1$ algebra. This will hopefully provide a new perspective on how the quantum group structure and associated quantum spacetime emerge in the DSSYK model.

\subsection{Information in de Sitter space}

One reason for taking a somewhat abstract approach in terms of operator algebras is that, a priori, it does not rely on any assumptions about the spacetime geometry. For this reason, the algebraic techniques seem particularly useful for studying the emergence of de Sitter space, see, for example, \cite{Chandrasekaran:2022cip, Gomez:2022eui, Seo:2022pqj, Gomez:2023upk, Gomez:2023wrq, Aguilar-Gutierrez:2023odp, Kudler-Flam:2024psh}. It was recently argued that the observable algebra of an observer in the static patch of de Sitter space is a type II$_1$ algebra \cite{Chandrasekaran:2022cip}, which is the algebraic setting that we have focused on in this paper. For this reason, it would be very interesting to come up with a natural inclusion of factors that arises in this context, perhaps by putting a Schwarzschild black hole in de Sitter space, and understand the interpretation of our algebraic recovery protocol in that case.

\section*{Acknowledgements}
We would like to thank Ramesh Chandra, Jan de Boer, Misha Isachenkov, Monica Kang, Bahman Najian, Antony Speranza, Jingxin Tu, Pim van den Heuvel and Herman Verlinde for useful discussions. JvdH is supported by the European Research Council under the European Unions Seventh Framework Programme (FP7/2007-2013), ERC Grant agreement ADG 834878. This work is supported by the Delta ITP consortium, a program of the Netherlands Organisation for Scientific Research (NWO) that is funded by the Dutch Ministry of Education, Culture and Science (OCW). 

\appendix

\section{Subfactors and the Jones index}

\label{app:A}

In this appendix, we have collected some results on subfactors and the Jones index for type II$_1$ von Neumann algebras. In particular, there is a precise definition of the Jones index. This appendix follows the lecture notes \cite{Speicher:2016}, so the interested reader can find a more in-depth discussion there that includes proofs of most of the statements.

Let $\mathcal{M}\subset \mathcal{B}(\mathcal{H})$ be a von Neumann algebra acting on some Hilbert space $\mathcal{H}$ with commutant $\mathcal{M}'$. We say that $\mathcal{M}$ is a factor if the center is trivial,  
\beq 
\mathcal{M}\cap\mathcal{M}' = \mathbb{C}1~.
\eeq
We will be mostly interested in type II$_1$ factors, which are characterized by being infinite-dimensional and having a unique trace $\mathrm{tr}:\mathcal{M}\to \mathbb{C}$. Let $\cM$ be such a type II$_1$ factor. Given the trace on $\cM$, there is a canonical representation which is obtained from the algebra through the Gelfand-Naimark-Segal (GNS) construction. The idea is to view $\cM$ as a linear space, and consider the action of the algebra $\cM$ on itself by left multiplication. The trace now induces an inner product on the algebra via the assignment
\beq 
\langle a, b \rangle \equiv \tr(b^*a)~.
\eeq
We will denote by $\mathcal{H}_{\cM}$ the Hilbert space completion of $\cM$ with respect to this inner product. 
\begin{definition}
The Hilbert space $\mathcal{H}_{\cM}$ is called the \emph{standard representation} of $\cM$.
\end{definition}
We represent the elements of $\mathcal{H}_{\cM}$ in terms of some distinguished vector $|\Psi\rangle$ by writing 
\beq 
|a\rangle \equiv a|\Psi\rangle~, \qquad a\in \cM~.
\eeq
We refer to $|\Psi\rangle$ as the tracial state, since the trace can now be expressed as the expectation value 
\beq 
\tr(x) = \langle \Psi| x |\Psi \rangle~.
\eeq

Note that the Hilbert space $\mathcal{H}_{\cM}$ is bigger than the usual fundamental representation that is used in the case of finite-dimensional matrix algebras. For concreteness, let $\cM=M_{n}(\mathbb{C})$ be the algebra of $n\times n$ matrices. The fundamental representation is now simply given by the vector space $\cH=\mathbb{C}^n$. However, a better representation for our purposes is the standard representation, which in this case is given by 
\beq \label{eq:GNStypeI}
\mathcal{H}_{\cM} = \mathbb{C}^{n^2}=\mathbb{C}^n\otimes \mathbb{C}^n~,
\eeq
which consists of two copies of the fundamental representation. The algebra of operators that acts on the Hilbert space \eqref{eq:GNStypeI} now decomposes according to 
\beq 
\mathcal{B}(\mathcal{H}_{\cM})=M_{n}(\mathbb{C})\otimes M_{n}(\mathbb{C})~.
\eeq
Given that the algebra $\mathcal{M}$ acts by left-multiplication, one can identify $\cM=M_{n}(\mathbb{C})\otimes 1$. On the other hand, the commutant algebra $\cM'$ that acts via right-multiplication can be represented as $\cM'=1\otimes M_{n}(\mathbb{C})$. Note that in the standard representation the algebras $\cM$ and $\cM'$ are the same size. 

\begin{definition}
A state $|\Omega\rangle \in \cH$ is called \emph{cyclic} for the algebra $\cM$ if one can generate the whole Hilbert space by acting with operators on the state, 
\beq 
\cH = \overline{\mathcal{M}|\Omega\rangle}~.
\eeq
A state $|\Omega\rangle \in \cH$ is called \emph{separating} for the algebra $\cM$ if for $a\in \cM$ we have 
\beq 
a|\Omega\rangle = 0 \quad \longrightarrow \quad a=0~.
\eeq
\end{definition}
The existence of a cyclic and separating vector tells us something about the relative sizes of the algebra $\cM$ and its commutant $\cM'$. In particular, having a cyclic vector means that $\cM$ is large enough to generate the whole Hilbert space $\cH$, while the existence of a separating vector for $\cM$ shows that $\cM'$ is sufficiently large to generate $\cH$. It is easy to check that the tracial state $|\Psi\rangle \in \mathcal{H}_{\cM}$ is both cyclic and separating for $\cM$. This follows from the definition of the GNS construction and the fact that $\tr$ is faithful. Therefore, in the standard representation both $\cM$ and its commutant $\cM'$ are large enough to generate the whole Hilbert space. 

Besides the standard representation, one can construct other representations\footnote{By a representation of $\cM$ we mean a $\cM$-module, that is, a Hilbert space $\cH$ with an action of $\cM$ via a unital $*$-homomorphism $\cM\to \mathcal{B}(\cH)$ that is continuous in an appropriate sense.} of the algebra $\cM$ as well. Let us give some examples. Starting from the standard representation $\mathcal{H}_{\cM}$, one can make a smaller module by using a projector $e\in \cM'$. We define
\beq 
\cH \equiv \mathcal{H}_{\cM} e~,
\eeq
which has a natural $\cM$-module structure. Note that we have written the projector on the right, since $\cM'$ acts on the standard representation via right multiplication. The size of the Hilbert space $\cH$ is $\tr(e)$ times the size of $\mathcal{H}_{\cM}$. Importantly, by reducing our standard representation we have lost the separating vector. The state $|\Psi e\rangle\in \mathcal{H}$ is no longer separating for the algebra $\cM$. To see this, we consider the projection operator 
\beq 
f \equiv J_{\cM}eJ_{\cM}\in \cM~,
\eeq
where $J_{\cM}$ is the modular conjugation operator. It is not difficult to show that $(1-f)|\Psi e\rangle =0$, and hence the state is annihilated by a non-zero operator $1-f\in \cM$, which contradicts the separating property. Note that the commutant of the algebra $\cM$ within the Hilbert space $\cH$ gets replaced by $e\mathcal{M}'e$. This is still a type II$_1$ factor, but somewhat smaller than the original algebra $\cM'$ by a factor of $\tr(e)$. This is the reason that the operators in the commutant cannot generate the full Hilbert space $\cH$.

We can also make larger modules by taking direct summands of the standard representation. We define
\beq 
\cH \equiv \mathcal{H}_{\cM}\oplus \cdots \oplus \mathcal{H}_{\cM}~,
\eeq
where we have taken $n$ copies of the Hilbert space $\mathcal{H}_{\cM}$. In this case, our state is no longer cyclic. Indeed, the state $|\Psi\rangle \oplus \cdots \oplus |\Psi\rangle$ when acted on with operators $a\in \cM$ only generates the diagonal part of $\cH$. The commutant is now replaced by $M_n(\mathbb{C})\otimes\cM'$ which is still a type II$_1$ factor, but in some sense $n$ times bigger. 

There is also an infinite amplification of $\mathcal{H}_{\cM}$ by taking an infinite number of copies in the direct summand construction. To ensure that the result is a Hilbert space we define it in terms of the square integrable sequences with values in the standard representation, 
\beq 
\cH \equiv \ell^2(\mathbb{N})\otimes \mathcal{H}_{\cM}~.
\eeq
Note that in this case the commutant is given by $ \mathcal{B}(\ell^2(\mathbb{N}))\otimes \cM'$, which is not a type II$_1$, but a type II$_{\infty}$ algebra, since there is no \emph{finite} trace on the algebra. 

Combining the above constructions, it is possible to define $\cM$-modules which are in some sense $\alpha$ times as big as the standard representation with $\alpha \in [0,\infty]$. The precise statement is that one can associate a number
\beq \label{eq:couplingconstant}
\dim_{\cM}\cH \in [0,\infty]~,
\eeq
called the coupling constant or $\cM$-dimension to the representation $\cH$ with the property that the standard representation has dimension 
\beq 
\dim_{\cM} \cH_{\cM} =1~.
\eeq
The number \eqref{eq:couplingconstant} should be thought of as the relative size of $\cH$ compared to the standard representation $\cH_{\cM}$. It characterizes the representations in the sense that $\dim_{\cM}\cH = \dim_{\cM}\mathcal{K}$ if and only if $\cH$ and $\mathcal{K}$ are unitarily equivalent as representations of $\cM$.

To define the coupling constant we will now make use of the following theorem:

\begin{theorem}
Let $\cM$ be a type II$_1$ factor and $\cH$ an $\cM$-module. Then, there exists an isometry
\beq 
v:\cH \to \ell^2(\mathbb{N})\otimes \mathcal{H}_{\cM}~, \qquad \mathrm{with} \qquad v a = (1\otimes a) v~, \quad a\in \cM~.
\eeq
Moreover, the operator 
\beq 
vv^* \in (1\otimes \cM)'=\mathcal{B}(\ell^2(\mathbb{N}))\otimes \cM'
\eeq
defines a projector, and given the trace $\tau \equiv \mathrm{Tr}\otimes \tr$ on the commutant algebra, it turns out that the number $\tau(vv^*)$ is independent of $v$.
\end{theorem}
A proof of this theorem can be found in \cite[Theorem 3.16]{Speicher:2016}. For now we will simply use the existence of the projector $vv^*$ to give the following definition:

\begin{definition}
We define the \emph{coupling constant} of $\cH$ by 
\beq 
\dim_{\mathcal{M}}\cH \equiv \tau(vv^*)\in [0,\infty]~.
\eeq
\end{definition}
Note that $\dim_{\mathcal{M}}\cH$ is equal to the trace of a projection operator $e=vv^*$ in a type II$_\infty$ algebra, so $\tau(e)$ can take on all values in $[0,\infty]$. Intuitively, the coupling constant compares the relative sizes of $\cM$ and $\cM'$. From the above definition, one can derive the following properties:
\begin{itemize}
\item[(i)] We have $\dim_{\cM}\cH = \dim_{\cM}\mathcal{K}$ if and only if $\cH \cong \mathcal{K}$.
\item[(ii)] If $\dim_{\cM}\cH = \infty$, the algebra $\cM'$ is a type II$_\infty$ factor. Otherwise, it is type II$_1$. 
\item[(iii)] For a countable set of modules $\mathcal{H}_i$ labeled by some index set $i\in I$ we have 
\beq 
\dim_{\cM}\left( \bigoplus_i \cH_i \right)= \sum_{i}\dim_{\cM}\cH_i~.
\eeq
\item[(iv)] For a projection $e \in \cM'$ we have
\beq \label{eq:idprojectore'}
\dim_{\cM}(\mathcal{H}e) = \tr_{\cM'}(e)\dim_{\cM}\mathcal{H} ~.
\eeq
\item[(v)] For a projection $e\in \cM$ we have 
\beq 
\dim_{e\cM e}(e\cH)=\frac{1}{\tr_{\cM}(e)}\dim_{\cM}\cH~.
\eeq
\item[(vi)] It holds that
\beq \label{eq:inversecoupling}
\dim_{\cM'}\cH = \frac{1}{\dim_{\cM}\cH}~.
\eeq
\end{itemize}

We are now ready to define the Jones index. 
\begin{definition}
Let $\cN\subset \cM$ be an inclusion of type II$_1$ factors. The \emph{Jones index} is defined as
\beq  \label{eq:indexdefinition}
[\cM:\cN]\equiv \dim_{\cN} \mathcal{H}_{\cM}~.
\eeq
\end{definition}
Roughly speaking, the index measures the relative size of $\cN$ in $\cM$. Note that the algebra $\cN$ acts on $\mathcal{H}_{\cM}$ as well, so we can consider $\mathcal{H}_{\cN}\subset \mathcal{H}_{\cM}$. We have the decomposition
\beq 
\mathcal{H}_{\cM}=\mathcal{H}_{\cN}\oplus \mathcal{H}_{\cN}^{\rm ortho}~,
\eeq
in terms of the orthogonal complement of the Hilbert space $\cH_{\cN}$. As a consequence, we have 
\beq 
\dim_{\cN}\mathcal{H}_{\cM}=\dim_{\cN}\mathcal{H}_{\cN}+\dim_{\cN}\mathcal{H}_{\cN}^{\rm ortho} = 1 + \dim_{\cN}\mathcal{H}_{\cN}^{\rm ortho}~,
\eeq
so it follows that $[\cM:\cN]\geq 1$. Observe that we have equality if and only if $\mathcal{H}_{\cM}=\mathcal{H}_{\cN}$, which implies that $\cM = \cN$. Therefore, $[\cM:\cN]=1$ if and only if $\cM=\cN$. We can also express the index in the following way:

\begin{proposition}
Let $\cN\subset \cM$ and $\cH$ a representation with $\dim_{\cN}\cH < \infty$. Then, we have
\beq \label{eq:indexA}
[\cM:\cN]=\frac{\dim_{\cN}\cH}{\dim_{\cM}\cH}~.
\eeq
\end{proposition}
The proof is rather straightforward and can be found in \cite[Proposition 4.3]{Speicher:2016}. Using the definition of the index \eqref{eq:indexdefinition} and the useful formula \eqref{eq:indexA}, one can derive some of the properties of the basic construction that we used in the main text, for example, equation \eqref{eq:indexpreserved} which expresses the fact that when the index is finite $[\cM:\cN]< \infty$ the basic extension algebra $\cM_1$ preservers the index,
\beq 
[\cM_1:\cM]=[\cM:\cN]~.
\eeq
First, note that since $[\cM:\cN]< \infty$, the commutant $\cN'$ is a type II$_1$ factor. As a consequence, the algebra $\cM_1=J_{\cM}\cN'J_{\cM}$ is also a type II$_1$ factor. One can now show that 
\beq 
[\cM_1:\cM] = \frac{\dim_{\cM}\cH_{\cM}}{\dim_{\cM_1}\cH_{\cM}}=\frac{1}{\dim_{\cM_1}\cH_{\cM}}=\frac{1}{\dim_{J\cN'J}\cH_{\cM}}=\frac{1}{\dim_{\cN'}\cH_{\cM}}~,
\eeq
which by virtue of \eqref{eq:inversecoupling} is equal to $[\cM:\cN]=\dim_{\cN}\cH_{\cM}$.

Another step in the recovery protocol that we did not derive in the main text is equation \eqref{eq:traceM1}, which states that for $a\in \cM$ we have
\beq \label{eq:traceM1e}
\mathrm{tr}_{\mathcal{M}_1}(e_{\mathcal{N}}a)=[\mathcal{M}:\mathcal{N}]^{-1}\,\mathrm{tr}_{\mathcal{M}}(a)~,
\eeq
where $e_{\cN}:\cH_{\cM}\to \cH_{\cN}$ is the Jones projector, and $\mathrm{tr}_{\cM_1}$ is the unique normalized trace on the type II$_1$ factor $\cM_1$. To prove this statement, it is useful to first show that 
\beq 
\mathrm{tr}_{\cM_1}(e_{\cN})=[\cM:\cN]^{-1}~.
\eeq
This follows from \eqref{eq:idprojectore'}. Indeed, since $e_{\cN}\in \cN'$ it holds that 
\beq 
1=\dim_{\cN}\cH_{\cN}=\dim_{\cN}e_{\cN}\cH_{\cM}= \mathrm{tr}_{\cN'}(e_{\cN})\dim_{\cN}\cH_{\cM}=\mathrm{tr}_{\cN'}(e_{\cN})\cdot[\cM:\cN]~.
\eeq
We can now use the fact that $\cM_1=J_{\cM}\cN'J_{\cM}$ and $J_{\cM}e_{\cN}J_{\cM}=e_{\cN}$, see \eqref{eq:M1} and \eqref{eq:JeJ} in the main text, to derive that 
\beq \label{eq:traceM1f}
\mathrm{tr}_{\cM_1}(e_{\cN})=\mathrm{tr}_{J_{\cM}\cN'J_{\cM}}(J_{\cM}e_{\cN}J_{\cM})=\mathrm{tr}_{\cN'}(e_{\cN})=[\cM:\cN]^{-1}~.
\eeq
This can be used to prove \eqref{eq:traceM1e}. Let us first take $b_1,b_2\in \cN$ and observe that, as $e_{\cN}\in \cN'$,
\beq 
\mathrm{tr}_{\cM_1}(b_1b_2e_{\cN})=\mathrm{tr}_{\cM_1}(b_1e_{\cN}b_2)=\mathrm{tr}_{\cM_1}(b_2b_1e_{\cN})~.
\eeq
This shows that the assignment $b\mapsto\mathrm{tr}_{\cM_1}(be_{\cN})$ satisfies the tracial property on the algebra $\cN$. By uniqueness of the trace on type II$_1$ factors, it follows that 
\beq \label{eq:lastequation}
\mathrm{tr}_{\cM_1}(be_{\cN}) = \alpha \cdot\mathrm{tr}_{\cN}(b) = \alpha \cdot\mathrm{tr}_{\cM}(b)~,
\eeq
for some constant $\alpha$, where we have used in the last step that the restriction of $\mathrm{tr}_{\cM}$ to $\cN$ is given $\mathrm{tr}_{\cN}$. Evaluating the above expression for $b=\mathds{1}$, we find that the constant in \eqref{eq:lastequation} is given by 
\beq 
\alpha = \mathrm{tr}_{\cM_1}(e_{\cN}) = [\cM:\cN]^{-1}~,
\eeq
as follows from \eqref{eq:traceM1f}. For operators $a\in \cM$ we now have 
\beq 
\mathrm{tr}_{\cM_1}(a e_{\cN}) = \mathrm{tr}_{\cM_1}(e_{\cN}a e_{\cN}) =\mathrm{tr}_{\cM_1}(\mathcal{E}(a)e_{\cN})=\alpha \cdot\mathrm{tr}_{\cM}(\mathcal{E}(a)) = \alpha \cdot\mathrm{tr}_{\cM}(a)~,
\eeq
where we have used some properties of the conditional expectation $\mathcal{E}$ associated to the Jones projector $e_{\cN}$, namely that it satisfies $e_{\cN}a e_{\cN}=\mathcal{E}(a)e_{\cN}$ and $\mathrm{tr}_{\cM}\circ \mathcal{E}=\mathrm{tr}_{\cM}$. We have therefore established \eqref{eq:traceM1e}.

\bibliographystyle{JHEP}
\bibliography{references}

\end{document}